\documentclass[aps,showpacs]{revtex4}

\usepackage{graphicx}
\usepackage{booktabs}
\usepackage{multirow}

\begin{document}

\title{Extreme fluctuations in noisy task-completion landscapes on scale-free networks}

\author{H. Guclu}
\affiliation{Center for Nonlinear Studies, Theoretical Division,
Los Alamos National Laboratory, MS-B258, Los Alamos, New Mexico, 87545, USA}

\author{G. Korniss}
\affiliation{Department of Physics, Applied Physics, and Astronomy,
Rensselaer Polytechnic Institute, 110 8$^{th}$ Street, Troy, New York, 12180-3590 USA}

\author{Z. Toroczkai}
\affiliation{Department of Physics, University of Notre
Dame, Notre Dame, Indiana, 46556 USA}


\begin{abstract}
We study the statistics and scaling of extreme fluctuations in
noisy task-completion landscapes, such as those emerging in synchronized
distributed-computing networks, or generic causally-constrained queuing networks, with
scale-free topology. In these networks the average size of the fluctuations
becomes finite (synchronized state) and the extreme fluctuations typically diverge only
logarithmically in the large system-size limit ensuring
synchronization in a practical sense. Provided that local fluctuations in the
network are short-tailed,
the statistics of the extremes are governed by the Gumbel distribution.
We present large-scale simulation results using the exact algorithmic rules,
supported by mean-field arguments based on a coarse-grained description.
\end{abstract}


\date{\today}

\maketitle


{\bf The understanding of the characteristics of fluctuations in
task-completion landscapes in distributed processing networks is
important from both fundamental and system-design viewpoints.
Here, we study the statistics and scaling of the extreme
fluctuations in synchronization landscapes of task-completion
networks with scale-free topology. These systems have a large
number of coupled components and the tasks performed on each
component (node) evolve according to the local synchronization
scheme. We consider short-tailed local stochastic task-increments,
motivated by certain distributed-computing algorithms implemented
on networks. In essence, in order to perform certain tasks,
processing nodes in the network must often wait for others, since
their assigned task may need the output of other nodes. Typically,
large fluctuations in these networks are to be avoided for
performance reasons. Understanding the statistics of the extreme
fluctuations in our model, will help us to better understand the
generic features of back-log formations and worst-case delays in
networked processing systems.
We find that the average size of the fluctuations in the
associated landscape on scale-free networks becomes finite and largest fluctuations
diverge only logarithmically in the large system-size limit. This
weak divergence ensures an autonomously-synchronized, near-uniform
progress in the distributed processing network. The statistics of
the maximum fluctuations on the landscape is governed by the
Gumbel distribution.}

\begin{section}{Introduction}

Many artificial and natural systems can be described by models of
complex networks
\cite{BOCCALETTI06,NEWMAN03,DOROGOVTSEV02,ALBERT02,BOLLOBAS01}.
The ubiquity of complex networks has led to a dramatic increase in
the study of the structure of these systems. Recent research on
networks has shifted the focus from the structural (topological)
analysis to the study of processes (dynamics) in these complex
interconnected systems. The main problem addressed in these
studies is how the underlying network topology influences the
collective behavior of the system.

Synchronization is a good example for processes in networks and it
is also a fundamental problem in natural and artificial coupled
multi-component systems \cite{Strogatz_review}. Since the
introduction of small-world (SW) networks \cite{WATTS98,WATTS99},
it has been established that such networks can facilitate
autonomous synchronization \cite{BARAHONA02,WANG02,HONG02_1}.
Synchronization in the context of coupled nonlinear dynamical
systems such as chaotic oscillators has been also studied in
scale-free networks
\cite{WANG02_2,LAI03,MOTTER05a,MOTTER05b,ZHOU06a,ZHOU06b}. In
these studies the ratio of the largest and the smallest non-zero
eigenvalues of the network Laplacian (in the linearized problem)
has been used as a measure for ``desynchronization'', i.e.,
smaller ratios corresponding to better synchronizability.

Another synchronization problem emerges in the context of parallel
discrete-event simulations
\cite{LUBA,KORN00_PRL,SLOOT01,KORNISS03a,KIRKPATRICK03,KOLA_REV_2005}.
Nodes must frequently ``synchronize" with their neighbors (on a given network) to ensure
causality in the underlying simulated dynamics. The local synchronizations, however, can
introduce correlations in the resulting task-completion landscape,
leading to strongly non-uniform progress at the individual
processing nodes. The above is a prototypical example for
\textit{task-completion landscapes} in causally-constrained
queuing networks \cite{TORO03}. Analogous questions can also be
posed in supply-chain networks based on electronic transactions
\cite{NAGURNEY05} etc. The basic task-completion model has been
considered for regular (in 1D
\cite{KORN00_PRL,KORNISS01_MRS,KORN_ACM} and 2D
\cite{GUCLU_PRE,GUCLU_THESIS}), small-world \cite{KORNISS03a} and
scale-free (SF) networks \cite{TORO03,GUCLU_THESIS,KORNISS06}.
The extreme fluctuations in SW networks also have been studied
\cite{GUCLU_CNLS,GUCLU05_FNL} previously. Here, we provide a
detailed account and new results on the extreme fluctuations in
task-completion networks with SF structure. Through our
study, one also gains some insight into the effects of SF
interaction topologies on the suppression of critical fluctuations
in interacting system.

The field of extremes has attracted the attention of engineers,
scientists, mathematicians and statisticians for many years. From
an engineering point of view, physical structures need to be
designed such that special attention is paid to properties under
extreme conditions requiring an understanding of the statistics of
extremes (minima and maxima) in addition to average values.
\cite{FT,GUMBEL,GALAMBOS}. For example, in designing a dam,
engineers, in addition to being interested in average flood, which
gives the total amount of water to be stored, are also interested
in the maximum flood, the maximum earthquake intensity or the
minimum strength of the concrete used in building the dam
\cite{CASTILLO05}. Extreme-value theory is unique as a statistical
discipline in that it develops techniques and models for
describing the unusual rather than the usual \cite{COLES01}.
Similarly, in networked processing systems, in addition to the
average ``load" or progress, knowing the typical size and the
distribution of the extreme fluctuations is of great importance,
since failures and delays are triggered by extreme events
occurring on an individual node or link.

The relationship between extremal statistics and universal
fluctuations of {\em global} order parameters in correlated
systems has been the subject of recent intense research
\cite{BRAMWELL_NATURE,BRAMWELL_PRL,WATKINS02,
BRAMWELL_PRL2,BRAMWELL_PRE,GOLD,ANTAL2001,ANTAL2002,
DAHLSTEDT01,CHAPMAN,GYORGY03,BOUCHAUD, BALDASSARRI02,BERTIN}.
Closer to our interest, universal distributions for explicitly the
{\em extreme} ``height" fluctuations have been studied for fluctuating surfaces
\cite{SHAPIR,MAJUMDAR04,MAJUMDAR05,MAJUMDAR05_2,MAJUMDAR06,Gyorgyi_2006},
in particular for the Kardar-Parisi-Zhang (KPZ) surface growth
model \cite{KARD86} in one dimension \cite{MAJUMDAR04,MAJUMDAR05}.
It turns out that the basic task-completion landscape (emerging in
certain synchronized distributed computing schemes), on regular lattices,
belongs precisely to the KPZ universality class
\cite{KORN00_PRL,GUCLU_PRE}. In this paper, we address the suppression of
the extreme fluctuations of the local order parameter (local
progress) in scale-free (SF) noisy task-completion networks.

This paper is organized as follows. In Section~II we give a brief
review on the extreme-value statistics of independent and
identically-distributed random variables, with some further
details for exponential-like random variables. Section~III
describes our prototypical model for task-completion systems and
provides a mathematical framework to analyze the evolution of its
progress landscape. We present our results in Section~IV and
finish the paper with conclusions and summary in Section~V.

\end{section}

\begin{section}{Extreme-Value Statistics}

Extreme-value theory deals with stochastic behavior of the maxima
and minima of random variables. Let us first focus on independent
and identically distributed (iid) random variables. The
distributional properties of the extremes are determined by the
tails of the underlying individual distributions. By definition
extreme values are scarce implying an extrapolation from observed
levels to unobserved levels, and extreme value theory provides a
class of models to enable such extrapolation \cite{COLES01}.

Historically, work on extreme value problems may be traced back to
as early as 1700s when Bernoulli discussed the mean largest
distance from the origin given some points lying at random on a
straight line of a fixed length (See \cite{GUMBEL}). Theoretical
developments of extreme-value statistics in 1920s
\cite{BORTKIEWICZ22, MISES23, DODD23, FRECHET27, FT} were followed
by research dealing with practical applications in radioactive
emissions \cite{GUMBEL37}, flood analysis \cite{GUMBEL41},
strength of materials \cite{WEIBULL39}, seismic analysis
\cite{NORDQUIST45}, rainfall analysis \cite{POTTER49} etc. In
terms of applications, Gumbel \cite{GUMBEL}, made several
contributions to extreme value analysis and called the attention
of engineers and statisticians to applications of the
extreme-value theory. Here, we review \cite{KOTZ00} the basics on
the statistics of the {\em maximum} of $N$ iid random variables.

Let $X$ be a random variable with probability density function
(pdf) $f(x)$ and cumulative distribution function (cdf) $F(x)$
(the probability that the individual stochastic variable is less
than $x$). $f(x)$ and $F(x)$ are also referred to as the parent
distributions. Let $\{X_i\}_{i=1}^N$ be an iid sample drawn from
$f(x)$. Then the joint pdf can be written as

\begin{equation}
f(X_1,X_2,...,X_N) = \prod_{i=1}^{N} f(X_i)
\label{joint_pdf}
\;,
\end{equation}
and their joint cdf as
\begin{equation}
F(X_1,X_2,...,X_N) = \prod_{i=1}^{N} F(X_i)
\label{joint_cdf}
\;.
\end{equation}
Then the cdf of the maximum order statistic,
$X_M=\max\{X_1,X_2,...,X_N\}$, for iid random variables, can be
written as
\begin{eqnarray}
F_M(x) &=& {\rm Pr}(X_M\leq x) \nonumber \\
&=& {\rm Pr}(\text{max}\{X_1,X_2,...,X_N\}\leq x) \nonumber \\
&=& {\rm Pr}(X_1 \leq x; X_2 \leq x; \ldots; X_N \leq x;) \nonumber \\
&=& \prod_{i=1}^{N}{\rm Pr}(X_i\leq x) = \prod_{i=1}^{N} F(x) = [F(x)]^N
\label{joint_pdf_max}
\;.
\end{eqnarray}
The pdf of $X_M$ can be calculated by differentiating the equation
above with respect to $x$, yielding
\begin{equation}
f_M(x) = Nf(x)[F(x)]^{N-1}
\label{pdf_cdf}
\;.
\end{equation}

In many situations, extreme value analysis is built on a
sequence of data that is block (sample) maxima or minima. A
traditional discussion on the mean of the sample is based on the
central limit theorem and it forms the basis for statistical inference theory
\cite{BEIRLANT04}. The central limit theorem deals
with the statistics of the  sum $S_N$$=$$X_1+X_2+...+X_N$ (proportional to the
arithmetic average) and provides the constants
$a_N$ and $b_N$$>$$0$ such that $Y_N$$=$$(S_N-a_N)/b_N$ tends in
distribution to a non-degenerate distribution. In the case when $X$ has finite
variance, this distribution is the normal distribution. However, when the
underlying distribution has a slowly decaying (or heavy) tail, some
other stable distributions are attained instead of normal
distribution \cite{BOUCHAUD00}. Specifically, power-law
distributions with infinite variance will yield non-normal limits
for the average: the extremes produced by such a sample will
``corrupt" the average so that an asymptotic behavior different
from the normal behavior is obtained \cite{BEIRLANT04}.

As the sample size $N$ goes to infinity, it is clear that for any fixed value of $x$
the distribution of the maxima becomes
\begin{equation}
\lim_{N\rightarrow \infty} F_{M}(x) =
\left\{
\begin{array}{ll}
1 & \text{if}~F(x) = 1\\
0 & \text{if}~F(x) < 1\\
\end{array}
\right. \;,
\label{limit_fM}
\end{equation}
which is a degenerate distribution (it takes the values $0$ and
$1$ only). If there is a limiting distribution, one has to obtain
it in terms of a sequence of transformed (or reduced) variable,
such as $(X_M-a_N)/b_N$ where $a_N$ and $b_N(>0)$ may depend on
$N$ but not $x$. The main mathematical challenge here is finding
the sequence of numbers ($a_N$ and $b_N$) such that for all real
values of $x$ (at which the limit is continuous) the limit goes to
a non-degenerate distribution.
\begin{equation}
Pr(\frac{X_M-a_N}{b_N}\leq x) \rightarrow
G_M(x)~\text{as}~N\rightarrow \infty
\label{limit_gx} \;.
\end{equation}

The problem is twofold: (i) finding all possible (non-degenerate)
distributions G that can appear as a limit in
Eq.~(\ref{limit_gx}); (ii) characterizing the distributions F for
which there exist sequences $\{a_N\}$ and $\{b_N\}$ such that
Eq.~(\ref{limit_gx}) holds for any such specific limit
distribution \cite{BEIRLANT04}. The first problem is the
\textit{(extremal) limit problem} and has been solved in
\cite{FRECHET27,FT,GNEDENKO43} and later revived in \cite{HAAN70}.
The second part of the problem is called the \textit{domain of
attraction problem}. Under the transformation through $a_N$ and
$b_N$ the extreme types theorem states that the non-degenerate
distribution $G_M$ belongs to one of the following families
\begin{equation}
G_M(x) =
\left\{
\begin{array}{ll}
\exp[-\exp(\frac{\lambda-x}{\sigma})], & -\infty < x < \infty\\
\end{array}
\right. \;,
\end{equation}
\begin{equation}
G_M(x) =
\left\{
\begin{array}{ll}
0, & \text{if}~x<\lambda\\
\exp[-(\frac{\sigma}{x-\lambda})^\omega], & \text{otherwise}\\
\end{array}
\right. \;,
\end{equation}
\begin{equation}
G_M(x) =
\left\{
\begin{array}{ll}
\exp[-(\frac{\lambda-x}{\sigma})^\omega], & \text{if}~x< \lambda\\
1, & \text{otherwise}\\
\end{array}
\right. \;.
\end{equation}
Collectively, these three classes of distributions are widely known as
Gumbel, Fr\'echet and Weibull distributions, respectively. Each family has a
location and a scale parameter, $\lambda$ and $\sigma$, respectively;
additionally, the Fr\'echet and Weibull families have a shape parameter $\omega$.
The above theorem implies that when $X_M$ can be stabilized with
suitable sequences $a_N$ and $b_N$, the corresponding normalized
variable $X_M^*$$=$$(X_M$$-$$a_N)$$/$$b_N$ has a limiting
distribution that must be one of the three types of extreme value distribution.
The remarkable feature of this result is that the three types of
extreme-value distributions are the only possible limits for the
distributions of the $X_M^*$, regardless of the parent
distribution $F$ for the population. In this sense, the theorem is
an extreme-value analog of the central limit theorem \cite{COLES01}.
\begin{table}
\centering
\caption{Domain of attractions of the most common distributions
for the maximum of iid random variables.}
\begin{tabular}{l l l}
\\[1ex]
\hline\hline
\toprule
Distribution & Domain\\
\hline
\midrule
Normal      & Gumbel \\
Exponential & Gumbel \\
Lognormal   & Gumbel \\
Gamma       & Gumbel \\
Uniform     & Weibull \\
Pareto      & Fr\'echet \\
\bottomrule
\hline
\end{tabular}
\label{table_domains}
\end{table}

Now we briefly summarize, the basic properties for the maximal
values of $N$ independent stochastic variables
\cite{FT,GUMBEL,GALAMBOS,AB,BOUCHAUD} drawn from a generic {\em
exponential-like} individual pdf. We consider the case when the
parent complementary cdf (survival function),
$S(x)$$=$$1$$-$$F(x)$, (the probability that the individual
stochastic variable is larger than $x$) decays faster than any
power law in the tail, i.e., exhibits an exponential-like tail in
the large-$x$ limit. (Note that in this case the corresponding
probability density function displays the same exponential-like
asymptotic tail behavior.)  Using Eq.~(\ref{joint_pdf_max}) the
cumulative distribution $F_M(x)$ for the largest of the $N$ events
(the probability that the maximum value is less than $x$) can be
approximated as \cite{BOUCHAUD,BOUCHAUD00,AB}
\begin{equation}
F_M(x) = [F(x)]^N = [1-S(x)]^N = e^{N\ln[1-S(x)]} \simeq e^{-NS(x)} \;,
\end{equation}
where one typically assumes that the dominant contribution to the
statistics of the maximum comes from the tail of the individual
distribution. Now we assume $S(x)\simeq e^{-cx^{\delta}}$ for
large $x$ values, where $c$ is a constant and $\delta$ characterizes
the exponential-like tail. This yields
\begin{equation}
F_M(x) \simeq e^{-e^{-cx^{\delta}+\ln(N)}}.
\label{raw_extreme}
\end{equation}

The extreme-value limit theorem implies that there exists a
sequence of scaled variables $\tilde{x}=(x-a_{N})/b_{N}$, such
that in the limit of $N$$\to$$\infty$, the extreme-value
probability distribution for $\tilde{x}$ asymptotically approaches
the standard form of the Gumbel (also known as Fisher-Tippet Type I) distribution
\cite{FT,GUMBEL}:
\begin{equation}
G_M(\tilde{x}) \simeq e^{-e^{-\tilde{x}}} \;,
\label{gumbel}
\end{equation}
with the corresponding pdf
\begin{equation}
g_M(\tilde{x}) \simeq e^{-\tilde{x}-e^{-\tilde{x}}} \;,
\label{gumbel_pdf}
\end{equation}
with mean $\langle\tilde{x}\rangle$$=$$\gamma$
($\gamma$$=$$0.577\ldots$ being the Euler constant) and variance
$\sigma_{\tilde{x}}^{2}$$=$$\langle\tilde{x}^{2}\rangle$$-$$\langle\tilde{x}\rangle^{2}$$=$$\pi^2/6$.
From Eqs.~(\ref{raw_extreme}) and (\ref{gumbel}), one can deduce
\cite{AB} that to leading order, the scaling coefficients must be
$a_{N}$$=$$[\ln(N)/c]^{1/\delta}$ and $b_{N}$$=$$(\delta
c)^{-1}[\ln(N)/c]^{(1/\delta)-1}$. Note that for
$\delta$$\neq$$1$, while the convergence to
Eq.~(\ref{raw_extreme}) is fast, the convergence for the
appropriately scaled variable to the universal Gumbel distribution
in Eq.~(\ref{gumbel}) is {\em extremely} slow \cite{FT,AB}. The
average value of the largest of the $N$ iid variables with an exponential-like tails then
scales as
\begin{equation}
\langle x_{\max}\rangle = a_N + b_N\gamma \simeq [\ln(N)/c]^{1/\delta}
\label{mean}
\end{equation}
(up to ${\cal O}(\frac{1}{\ln(N)})$ corrections) in the asymptotic
large-$N$ limit. When comparing with experimental or simulation data, instead of Eq.~(\ref{gumbel}),
it is often convenient to use the Gumbel distribution scaled to zero mean and unit variance, yielding
\begin{equation}
G_M(y) = e^{-e^{-(ay+\gamma)}}
\;,
\label{gumbel_scaled}
\end{equation}
where $a$$=$$\pi/\sqrt{6}$ and $\gamma$ is the Euler constant. In particular, the corresponding Gumbel
pdf becomes
\begin{equation}
g_M(y) = ae^{-(ay+\gamma)-e^{-(ay+\gamma)}}
\;.
\label{gumbel_pdf_scaled}
\end{equation}

The mathematical arguments in obtaining the limit distributions
above assume an underlying process consisting of a sequence of
independent and identically distributed random variables. The most
natural application of a sequence of independent random variables
is to a stationary series. For some physical processes,
stationarity is a reasonable assumption and corresponds to a
series whose variables may be mutually dependent but whose
stochastic properties are homogeneous in time. There, the main
problem is finding the form of stationarity in terms of the range
of dependence. Then one attempts to find the timescale of the
series in which extreme events are almost independent. This is a
strong assumption, but there are a number of empirical stationary
series satisfying this property \cite{Eichner_2006}. Then,
eliminating the long-range dependence of extremes provides an opportunity
to consider only the effect of short-range (or weak) dependence by
using some rigorous \cite{GALAMBOS,Berman_1964} or
heuristic arguments leading to simple quantification in
terms of the standard extreme-value limits.

In this paper we will not discuss in detail the basic formulation
and treatment of the extreme limit distributions of dependent
random variables. Detailed work on limit distributions and
conditions required for different kinds of sequences such as
Markov, m-dependent, moving average, normal sequences etc. can be
found in the literature \cite{COLES01,CASTILLO05}. Most of the
research focused on weakly correlated random variables
\cite{GALAMBOS,Berman_1964} and only recent results have become
available on the statistical properties of the extremes of {\em
strongly correlated} variables
\cite{MAJUMDAR04,MAJUMDAR05,MAJUMDAR05_2,MAJUMDAR06,SHAPIR,Gyorgyi_2006}.
Traditional approaches, based on effectively uncorrelated
variables, immediately break down. Only recently, Majumdar and
Comtet \cite{MAJUMDAR04,MAJUMDAR05} obtained {\em analytic}
results for the distribution of extreme-height fluctuations in the
simplest strongly correlated fluctuating landscape: the steady
state of the one dimensional surface-growth EW/KPZ model.

As it becomes apparent, in light of recent results on SW
\cite{GUCLU04_PRE,GUCLU05_FNL} and our new results on SF
networks presented here (Sec. IV), implementing our stochastic nonlinear rules
for the task-completion model on a complex interaction topology,
in effect, ``eliminates" the complexity of the task-completion
landscape. While in low dimensions and regular topologies
fluctuations are strongly correlated and ``critical", in that they
are controlled by a diverging correlation length, on complex
networks, correlations become weak (or mean-field like), and one expects
the extreme fluctuations in the task-completion landscape to be
effectively governed by the traditional extreme-value limit
distributions for {\em independent} variables.

\end{section}

\begin{section}{Task-Completion Networks}

\begin{subsection}{The Model and Its Coarse-grained Description}

Consider an arbitrary network in which the nodes interact
through the links. The nodes are assumed to be task processing units,
such as computers or manufacturing devices.
Each node has completed an amount of tasks $\tau_i$  and these
together (at all nodes) constitute the {\em task-completion landscape}
$\{\tau_{i}(t)\}_{i=1}^{N}$.
Here $t$ is the discrete number of
parallel steps executed by all nodes, which is proportional to the
real time and $N$ is the number of nodes. At each
parallel step $t$, only certain nodes can receive additional tasks and when
that happens we say that an {\em update} happened at those nodes.
In this particular model the nodes that are allowed to update at a given step
are those whose completed task amount is
 not greater than the tasks at their
neighbors. We also choose the amount of new tasks arriving at a node to be
a random variable distributed according to an exponential distribution (Poisson asynchrony).
An example of a system being described by this model is a parallel computer
simulating short-range correlated discrete events in continuous time with
a Poisson inter-arrival time distribution between the events (independent
events) \cite{LUBA,KORN00_PRL,SLOOT01,KORNISS03a,KIRKPATRICK03,KOLA_REV_2005}.
Thus, denoting the
neighborhood of the node $i$ by $S_{i}$, if $\tau_{i}(t)\leq\min_{j\in
S_{i}}\{\tau_{j}(t)\}$, the node $i$ completes some
additional exponentially distributed random amount of task.
Otherwise, it idles.
In its simplest form the evolution equation for the amount of
task completed at the node $i$ can be written
\begin{equation}
\tau_i(t+1) = \tau_i(t) + \eta_{i}(t)\prod_{j\in S(i)}
\Theta\left(\tau_{j}(t)-\tau_i(t)\right)\;,
\label{revolution}
\end{equation}
where $\tau_i$ is the local field variable (amount of task
completed) at node $i$ at time $t$, $\eta_{i}(t)$ are identical,
exponentially distributed random variables with unit mean,
delta-correlated in space and time (the new task amount), and $\Theta(...)$ is the
Heaviside step-function. Despite its simplicity, this rule
preserves unaltered the asynchronous causal dynamics of the
underlying task-completion system \cite{LUBA,KORN00_PRL}.

While the dynamics above Eq.~(\ref{revolution}) is motivated by
the precise algorithmic rule in parallel discrete-event
simulations (PDES) \cite{LUBA,KORN00_PRL,KORNISS03a}, it also has
broader applications in ``causally-connected''
stochastic multi-component systems \cite{TORO03}:
The ``neighborhood'' local minima
rule [Eq.~(\ref{revolution})] is an essential ingredient of
generic {\em causally-constrained queuing networks}
\cite{KIRKPATRICK03}. In order to perform certain tasks,
processing nodes in the queuing/processing network often must wait for others,
since their assigned task may need the output of other
nodes.
Examples include manufacturing supply chains and various
e-commerce based services facilitated by interconnected servers
\cite{NAGURNEY05,Dong2003}. Understanding the statistics of the
extreme fluctuations in our model, will help us to better
understand the generic features of {\em back-log formations} and
{\em worst-case delays} in networked processing systems.

While the local synchronization rule gives rise to
strongly non-linear effective interactions between the nodes, we can gain some
insight by considering a linearized version of the corresponding
coarse-grained equations. As we have shown
\cite{KORN00_PRL,KORNISS03a,GUCLU_PRE}, neglecting non-linear
effects, the dynamics of the exact model in Eq.~(\ref{revolution})
can be {\em effectively} captured by the Edwards-Wilkinson (EW)
process \cite{EW} {\em on} the respective network. The EW process
on the network is a prototypical synchronization problem in a
noisy environment, where interaction between the nodes is
facilitated by simple relaxation
\cite{KORNISS03a,KORN00_PRL,KOZMA03,KORN_PLA_swrn} through the
links:
\begin{equation}
\partial_t \tau_i=-\sum_{j=1}^N C_{ij}(\tau_i-\tau_j) + \eta_i(t) \;.
\label{evolution}
\end{equation}
Here, $\langle \eta_i(t)\eta_j(t')\rangle$$=$$2\delta_{ij}\delta(t-t')$
and $C_{ij}$ is the coupling matrix with $C_{ii}$$=$$0$. In this
work, for simplicity, we only consider the case of symmetric
couplings, $C_{ij}$$=$$C_{ji}$. By using $C_{ij}$ we can define
the network Laplacian,
\begin{equation}
 \Gamma_{ij} = \delta_{ij}C_i-C_{ij}
\;,
\label{laplacian}
\end{equation}
where $C_i$$=$$\sum_l C_{il}$ and rewrite Eq.~(\ref{evolution}) as
\begin{equation}
 \partial_t \tau_i = -\sum_{j=1}^N \Gamma_{ij}\tau_j + \eta_i(t)
\;.
\label{evolution2}
\end{equation}

The above mapping suggests that on
low-dimensional regular lattices task-completion landscapes will
exhibit kinetic roughening \cite{BARA95,HEALY95,KRUG97}. The
landscape width provides a sensitive measure for the average degree of
de-synchronization
\cite{KORNISS03a,KORN00_PRL}:
\begin{equation}
\langle w^2(N,t) \rangle =
\left\langle\frac{1}{N}\sum_{i=1}^{N}[\tau_i(t)-\bar{\tau}(t)]^2\right\rangle
\;,
\label{width}
\end{equation}
where $\langle ... \rangle$ denotes an ensemble average over the noise
and $\bar{\tau}(t)$$=$$(1/N)\sum_{i=1}^N \tau_i(t)$ is the mean value of the local
task at time $t$. In addition to the width, we will study the scaling
behavior of the average of the {\em largest fluctuations} above the mean
in the steady-state regime
\begin{equation}
 \langle \Delta_{\max} \rangle = \langle \tau_{\max}(t)-\bar\tau(t) \rangle
\;,
\label{delta_max}
\end{equation}
where $\tau_{\max}(t)=\max\{\tau_1(t),\tau_2(t),\dots,\tau_N(t)\}$.

Since we use the formalism and terminology of non-equilibrium
surface growth phenomena, we briefly review scaling concepts for
self-affine or rough surfaces,  on regular spatial
lattices. The scaling behavior of the width, $\langle w^{2}(N,t)
\rangle$, alone typically captures and identifies the universality
class of the non-equilibrium growth process
\cite{BARA95,HEALY95,KRUG97}. In a finite system the width
initially grows as $\left\langle
w^2(N,t)\right\rangle$$\sim$$t^{2\beta}$, and after a system-size
dependent cross-over time $t_{\times}$$\sim$${N}^{z}$, it reaches
a steady-state value $\left\langle
w^2(N,t)\right\rangle$$\sim$${N}^{2\alpha}$ for $t$$\gg$$t_x$. In
the relations above $\alpha$, $\beta$, and
$z$=$\frac{\alpha}{\beta}$ are called the roughness, the growth,
and the dynamic exponent, respectively. In this work, we will only consider
the steady-state properties of the associated task-completion landscapes.

\end{subsection}

\begin{subsection}{Previous Work: Extreme-Fluctuations in Regular and SW Networks}

In one dimension on a regular lattice, with the relevant
non-linearities taken into account, we have shown
\cite{KORN00_PRL,TORO00} that the evolution of the task-completion
landscape Eq.~(\ref{revolution}) belongs to the
the Kardar-Parisi-Zhang (KPZ) \cite{KARD86} universality class \cite{BARA95}.
Indeed, when simulating the precise rule given by Eq.~(\ref{revolution}), the
evolution of the associated task-completion landscape exhibits KPZ-like kinetic
roughening \cite{KORN00_PRL,GUCLU_PRE}.
Further, in the steady-state, fluctuations are governed by the
Edwards-Wilkinson Hamiltonian \cite{EDWARDS82}.

In regular networks, the task-completion landscape is rough
\cite{KORN00_PRL,GUCLU_PRE} (de-synchronized state), i.e., it is dominated by
large-amplitude {\em long-wavelength} fluctuations. The extreme
local fluctuations emerge through these long-wavelength modes and,
in one-dimensional regular networks, the extreme and average
fluctuations follow the same power-law divergence with the system
size
\cite{SHAPIR,KORN_ACM,GUCLU_CNLS,GUCLU05_FNL,MAJUMDAR04,MAJUMDAR05,Gyorgyi_2006}
\begin{equation}
 \langle \Delta_{\max} \rangle \sim \sqrt{\langle w^2 \rangle} \sim N^\alpha \sim N^{1/2}
\;,
\label{delta_max_bcs}
\end{equation}
where $\alpha$$=$$1/2$ is the roughness exponent for the KPZ/EW
surfaces \cite{BARA95}. On this regular lattice, the
average size of the largest fluctuations {\em below} the mean
($\langle \Delta_{\min}\rangle = \langle
\bar\tau-\tau_{\min}\rangle$) and the maximum spread ($\langle
\Delta_{\max-\min}\rangle = \langle \tau_{\max}-\tau_{\min}\rangle$)
follow the same scaling as the average maximum fluctuation with
the system size. The diverging width is related to an underlying
diverging length scale, the lateral correlation length, which
reaches the system size $N$ for a finite system. This divergent
width hinders the synchronization (near-uniform progress) in
low-dimensional regular task-completion networks
\cite{KORN_ACM,KORNISS01_MRS}.

The width distribution for the EW (or a steady-state
one-dimensional KPZ) class is characterized by a universal scaling
function, $\Phi(x)$, such that $P(w^2) = \langle w^2
\rangle^{-1}\Phi(w^2/\langle w^2 \rangle) = N^{-1}\Phi(w^2/N)$,
where $\Phi(x)$ can be obtained analytically for a number of
models including the EW class \cite{FOLTIN94}. The width
distribution for the task-completion system on a one-dimensional
network is shown in Fig.~\ref{fig_dist-regular}(a). Systems
with $N \geq  10^3$ show convincing data collapse onto this exact
scaling function. The convergence to the limit
distribution is very slow when compared to other microscopic
models, such as the single-step model \cite{BARA95,ANTAL96},
belonging to the same KPZ universality class.
\begin{figure*}
\begin{center}
\begin{tabular}{cc}
\includegraphics[keepaspectratio=true,angle=0,width=83mm]
{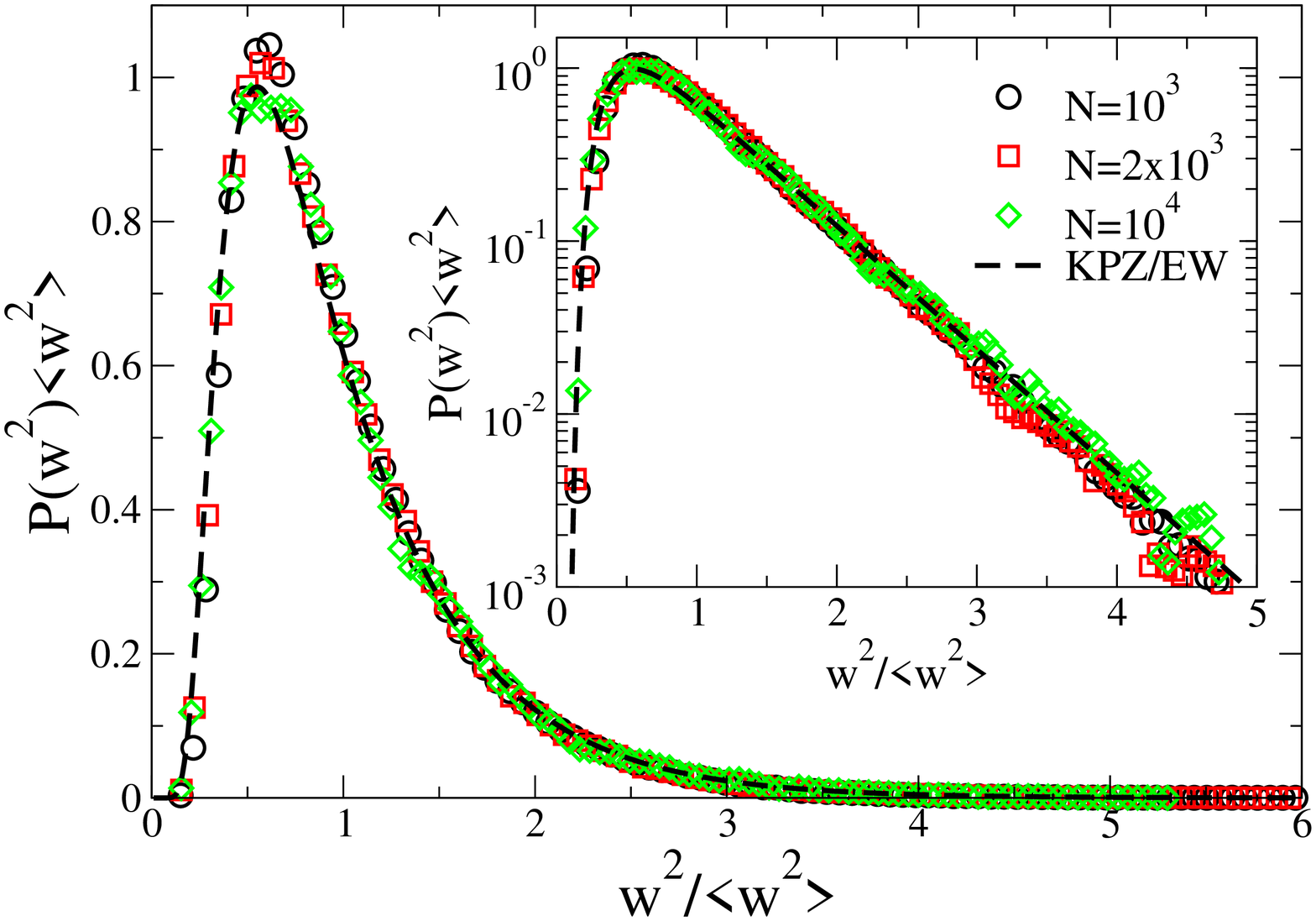} &
\includegraphics[keepaspectratio=true,angle=0,width=83mm]
{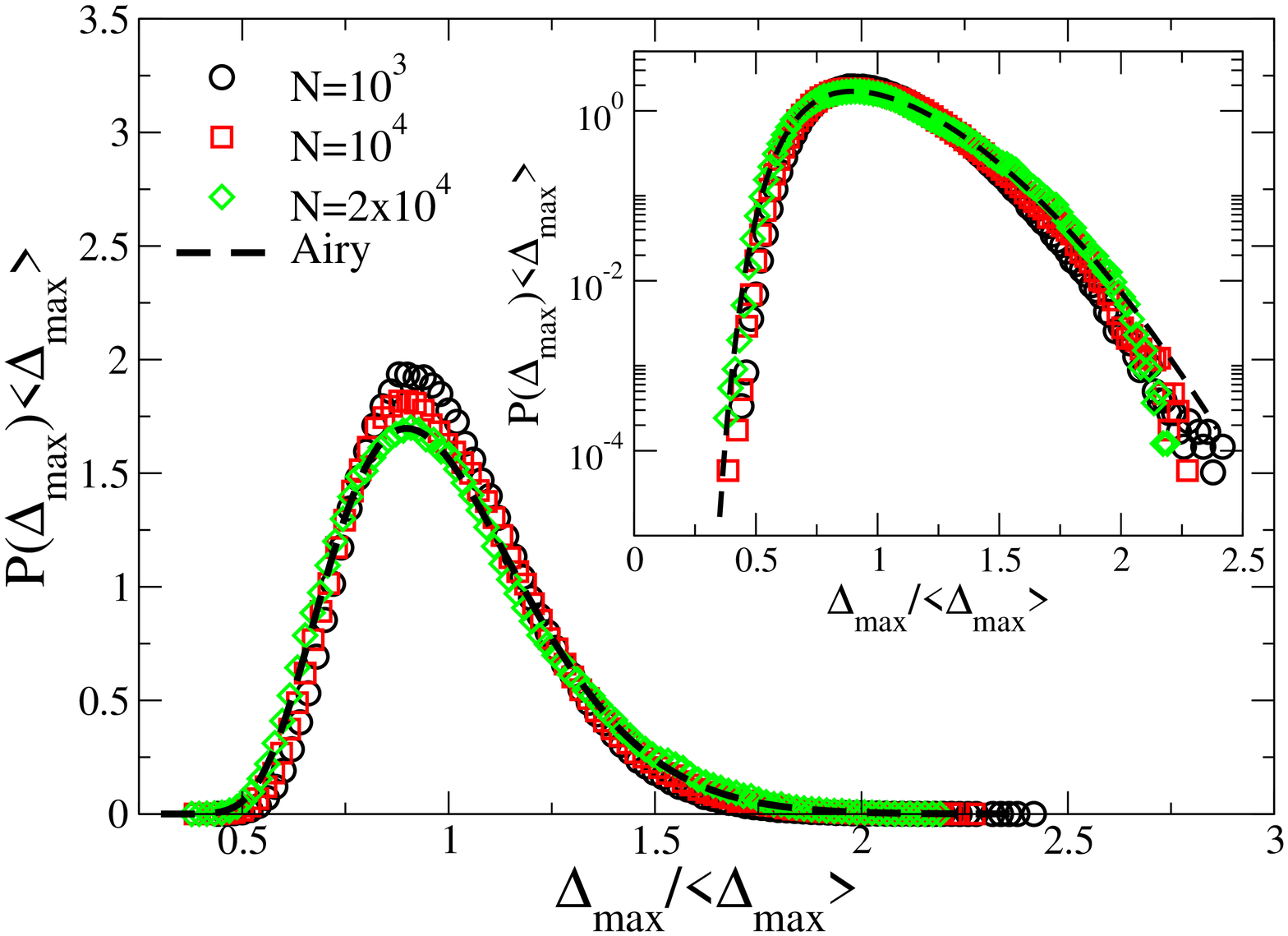}
\vspace{-0.3cm}
\\
\small{(a) Width distribution} & \small{(b) Maximum fluctuations distribution}
\end{tabular}
\end{center}
\vspace{-0.5cm}
\caption{(a) The scaled distribution of the width of task-completion
landscapes on a one-dimensional regular network. The inset is the same graph
in log-linear scale. The dashed curve is the scaled width distribution of
KPZ/EW surface \cite{FOLTIN94}.
(b) The scaled distribution of the maximum fluctuations in the same network.
The inset is in log-linear scale and the dashed curve is the appropriately scaled
Airy distribution function \cite{MAJUMDAR04,MAJUMDAR05}.}
\label{fig_dist-regular}
\end{figure*}

The extreme-value limit theorems summarized in the previous
section are valid only for independent (or short-range correlated)
random variables. Since the ``heights" (local progress) in
task-completion landscapes of regular networks are strongly correlated, the known
extreme-value limit theorems cannot be used. Some remarkable recent analytic
work yielded the distribution of the extreme
heights for the one-dimensional EW/KPZ steady-state surface
\cite{MAJUMDAR04,MAJUMDAR05,MAJUMDAR05_2,MAJUMDAR06}. Although the
local microscopic rule for the evolution of the task-completion
landscapes are different, they belong to the same EW/KPZ
universality class in one dimension \cite{BARA95}, and hence,
expected to exhibit the same universal distribution for the
extreme fluctuations. Equation~(\ref{delta_max_bcs}) suggests
that, similar to the width and its distribution, there is a single
scale governing the diverging extremes fluctuations, and hence,
the normalized probability density function of the maximum
relative fluctuations $\Delta_{\max}$ has a universal scaling form,
$P(\Delta_{\max})\sim N^{-\alpha} f(\Delta_{\max}/N^{\alpha})$. For
the 1D EW/KPZ surface with periodic boundary conditions
($\alpha$$=$$1/2$), by using path integral techniques, Majumdar
and Comtet \cite{MAJUMDAR04,MAJUMDAR05} found $f(x)$ to be the
so-called Airy distribution function. Our simulation results show
that the appropriately scaled maximum relative height
distributions are in agreement with the theoretical distribution
as can be seen in Fig.~\ref{fig_dist-regular}(b).

Recently, we have studied the extreme fluctuations in
task-completion landscapes on SW networks
\cite{GUCLU04_PRE,GUCLU05_FNL}. In particular, we considered two
SW-synchronized network models: in one case a small (variable)
density of random links were added on top of a one-dimensional
ring; in the other case, each node had exactly one random
(possibly long-range) connection, and the ``coupling strength"
(relative frequency of synchronization through the long-range
link) was varied. The basic findings were the same for both cases.
As a result of the non-zero density or non-zero strength of random
links, the correlation length $\xi$ becomes {\em finite} in such
networks (as opposed to the diverging correlation length on the
one-dimensional ring.) This important property is intimately
related to the emergence of an {\em effective} non-zero mass of
the corresponding network propagator (the inverse of the network
Laplacian) \cite{KOZMA03}. This is the fundamental effect of
extending the original dynamics to a SW network: it decouples the
fluctuations of the originally correlated landscape. Then, the
extreme-value limit theorems can be applied using the number of
independent blocks $N/\xi$ in the system \cite{BOUCHAUD,AB}. For
short tailed noise, the local individual task fluctuations also
exhibit short (exponential-like) tails,
$S(\Delta_i)\simeq\exp[-c(\Delta_i/w)^{\delta}]$, where
$\Delta_i$$=$$\tau_i$$-$$\bar{\tau}$ is the relative ``height"
measured from the mean at site $i$. (Note that the exponent
$\delta$ for the tail of the local relative height distribution
may differ from that of the noise as a result of the collective
(possibly non-linear) dynamics, but the exponential-like feature
does not change.) Then the (appropriately scaled) largest
fluctuations are governed by the Gumbel pdf [Eq.~(\ref{gumbel})],
and the average maximum relative height scales as
\begin{equation}
\langle \Delta_{\max}\rangle \simeq w
\left(\frac{\ln(N/\xi)}{c}\right)^{1/\delta} \simeq
\frac{w}{c^{1/\delta}}\left[\ln(N)\right]^{1/\delta} \;.
\label{mean_max}
\end{equation}
(Note, that both $w$ and $\xi$ approach their {\em finite}
asymptotic $N$-independent values for SW-coupled systems, and the
only $N$-dependent factor is $\ln(N)$ for large $N$ values.) In SW
synchronized systems with unbounded local variables driven by
exponential-like noise distribution (such as Gaussian), the
extremal fluctuations increase only {\em logarithmically} with the
number of nodes. This weak divergence, which one can regard as
marginal, ensures synchronization for practical purposes in
coupled multi-component systems.

Note that the exact ``microscopic'' dynamics
[Eq.~(\ref{revolution})] based on the task-completion rule, is
inherently non-linear, but the effects of the non-linearities only
give rise to a renormalized non-zero effective mass
\cite{KORNISS03a}. Thus, the synchronization dynamics is
effectively governed by EW relaxation in a SW, yielding a {\em
finite} correlation length and, consequently, the slow logarithmic
increase of the extreme fluctuations with the system size
[Eq.~(\ref{mean_max})]. Also, for the task-completion landscapes,
the local height distribution can be asymmetric with respect to
the mean, but the average size of the height fluctuations is, of
course, finite for both above and below the mean. This specific
characteristic simply yields different prefactors for the extreme
fluctuations [Eq.~(\ref{mean_max})] above and below the mean,
leaving the logarithmic scaling with $N$ unchanged.

Simulating the exact local task-completion rule
Eq.~(\ref{revolution}), we observed, that the local height
fluctuations exhibit simple exponential tails, hence $\delta=1$ and
the extremes scale as $\ln(N)$ with the number of nodes
\cite{GUCLU04_PRE,GUCLU05_FNL,GUCLU_CNLS}. The largest relative
deviations below the mean $\langle \Delta_{min} \rangle$, and the maximum spread
$\langle \Delta_{max-min} \rangle$ follows the same
logarithmic scaling with the system size $N$.

\end{subsection}

\end{section}

\begin{section}{Scale-Free Task-Completion Networks}

Recent studies show that many natural and artificial systems such
as the Internet, World Wide Web, scientific collaboration networks,
and e-mail networks have power-law degree (connectivity) distributions
\cite{BOCCALETTI06}, i.e., the probability
of having nodes with $k$ degrees is $P(k) \sim k^{-\gamma}$ where
$\gamma$ is usually between $2$ and $3$. These systems are commonly known as
power-law or scale-free networks since their degree distributions
are free of scale and follow power-law distributions over many orders of
magnitude.  Scale-free
networks have many interesting properties such as high tolerance to
random errors and attacks (yet low tolerance to attacks targeted at
hubs) \cite{ALB00}, high synchronizability
\cite{GUCLU_THESIS,TORO03,KORNISS06}, and resistance to congestion
\cite{TORO04_1,TORO04_2}.

In this paper, in part, we employed the \textit{Barab\'asi-Albert} (BA) model of network growth
inspired by the formation of World Wide Web \cite{BARABASI99}
to generate scale-free networks.
This model is based on two basic observations: growth and preferential attachment.
The basic idea is that the high-degree nodes attract links faster than
low-degree nodes. The network starts growing from $m_0$$=$$m+1$ nodes and at every
time step a new node with $m$ (``stubs'') possible links is added to the network.
The probability that the new node $j$ is connected to already existing node $i$ is
linearly proportional to the degree of the node $j$, i.e.,
${\rm Pr}(i\rightarrow j)$$=$$k_j/\sum_l k_l$. Once the given number of nodes $N$ is reached
in the network, the process is stopped. For the BA network, the degree
distribution is a power law in the asymptotic system size limit
($N$$\rightarrow$$\infty$, also called thermodynamic limit),
$P_{BA}(k) \simeq 2m^2/k^3$.(In obtaining this normalization,
one replaces the sum by the integral over the degree.)
Since every node has $m$ links initially, the network
at time $t$ will have $N$$=$$m_0+t$ nodes and $E$$=$$mt$ links, thus the average
degree $\langle k \rangle$$=$$2m$ for large enough $t$.
The special case of the model when $m=1$ creates a network without
any loops, i.e., the network becomes a tree with no clustering.

Another model we employed to generate scale-free networks and to
compare with the BA model is the \textit{configuration model} (CM). The
CM  \cite{BENDER78,MOLLOY95,MOLLOY98} was introduced as an
algorithm to generate random networks with a given degree
distribution. Although CM has been considered to generate
uncorrelated random networks, it was shown that it has
correlations, especially between the nodes with larger degrees
\cite{CATANZARO05,BOGUNA04}. In CM, the vertices of the graph are
assigned a sequence of degrees $\{k_i\}_{i=0}^{N}, m \leq
k_i \leq k_c$ from a desired distribution $P(k)$. (There is an
additional constraint that the $\sum_i k_i$ must be even.) Then,
pairs of nodes are chosen randomly and connected by undirected
edges. This model generates a network with the expected degree
distribution and no degree correlations; however, it allows
self-loops and multiple-connections when it is used as described
above. It was proven in Ref.~\cite{BOGUNA04} that the number of
multiple connections when the maximum degree is fixed to the
system size, i.e., $k_c$$=$$N$, scales with the system size $N$ as
$N^{3-\gamma}\ln N$. After this procedure we simply delete the
multiple connections and self-loops from the network which gives a
very marginal error in the degree distribution exponent. This  might also cause
that some negligible number of
nodes in the network to have degrees less than the fixed minimum
degree ($m$) value or even zero. One another characteristic of the
CM is that the network may not be a connected network for small
values of $m$ such as $1$ and even $2$, i.e., it has disconnected
clusters (or components). For high values of $m$, the network is
almost surely connected having one giant component including all
the nodes. The degree distribution of SF networks with degree
exponent $\gamma$ (for $k_c\gg m$) can be written as
\begin{equation}
P(k)\simeq(\gamma-1)m^{\gamma-1}k^{-\gamma}\;,
\label{P_k}
\end{equation}
where $m$ is the minimum degree in the network, and again, in obtaining the
above normalization, we replaced the sum by the integral over the degree.
Then the average and the minimum degree are then related through
$\langle k\rangle=m(\gamma-1)/(\gamma-2)$.

\begin{subsection}{Mean-Field and Exact Numerical Diagonalization Approaches
for the EW Process on SF Networks}

We, again, can gain some insight to the problem by first
considering the linearized effective equations of motion, i.e.,
the EW process [Eq.~(\ref{evolution})] on a SF network. In the
mean-field (MF) approximation, local task fluctuations {\em about
the mean} are decoupled and reach a stationary distribution with
variance \cite{KORNISS06}
\begin{equation}
\left\langle(\tau_i-\bar{\tau})^2\right\rangle\approx1/C_i \;.
\label{w_i}
\end{equation}
For identical (unweighted) couplings (with unit link strength
without loss of generality), $C_{ij}$ is simply the adjacency
matrix, hence, $C_i=\sum_{l}C_{il}=k_i$, i.e., the degree of node
$i$. Then, for the width, one can write
\begin{equation}
\langle w^2 \rangle =
\frac{1}{N}\sum_{i=1}^{N}\left\langle(\tau_i-\bar{\tau})^2\right\rangle
\approx \frac{1}{N}\sum_{i}\frac{1}{C_i}
= \frac{1}{N}\sum_{i}\frac{1}{k_i}
\approx \int_{m}^{\infty} dk \frac{P(k)}{k} \;,
\label{w2_MF}
\end{equation}
where using infinity as the upper limit in the above integral is
justified for SF networks as $N\to\infty$, since $\gamma>0$. Using
the degree distribution of SF networks given by Eq.~(\ref{P_k}), one
finally obtains the mean-field expression for the width
\begin{equation}
\langle w^2 \rangle \approx
\frac{1}{m} \frac{(\gamma-1)}{\gamma}
= \frac{1}{\langle k\rangle} \frac{(\gamma-1)^2}{\gamma(\gamma-2)}\;.
\label{w2_MF_final}
\end{equation}
The main message of the above result is that the width approaches
a {\em finite} value in the limit of $N\to\infty$, and for the
{\em linearized} problem, should scale as $\langle w^2 \rangle
\sim 1/m \sim 1/\langle k\rangle$.

Extracting the steady-state width from exact numerical
diagonalization \cite{KORN_PLA_swrn,KORNISS06} through
\begin{equation}
\langle w^2 \rangle = \frac{1}{N}\sum_{k=1}^{N-1} \frac{1}{\lambda_{k}} \;,
\label{w2_exact}
\end{equation}
where $\lambda_{k}$ are the {\em non-zero} eigenvalues of the
network Laplacian on the corresponding SF network, supports these
MF predictions [Fig.~\ref{fig_exact_w2}]. Except for the $m=1$ BA
network (when the network is a scale-free tree), finite-size effects are negligible as
the width approaches a finite value in the $N\to\infty$ limit
[Fig.~\ref{fig_exact_w2}(a)]. For $m=1$, the width weakly
(logarithmically) diverges with the system size [Fig.~\ref{fig_exact_w2}(a) inset].
A closer look at the spectrum reveals that the gap approaches a {\em non-zero}
value for $m>1$ as $N\to\infty$, while it slowly vanishes for the
$m=1$ BA network. As can be expected, Fig.~\ref{fig_exact_w2}(b)
(inset) indicates that MF scaling Eq.~(\ref{w2_MF_final}) for the
width works well for sufficiently large minimum (and average)
degree, $m\stackrel{>}{\sim}{\cal O}(10)$. Figure
~\ref{fig_exact_w2}(c) also shows results for the CM network
for two values of $\gamma$, and the corresponding MF result.
\begin{figure*}
\begin{center}
\begin{tabular}{ccc}
\includegraphics[keepaspectratio=true,angle=0,width=60mm]
{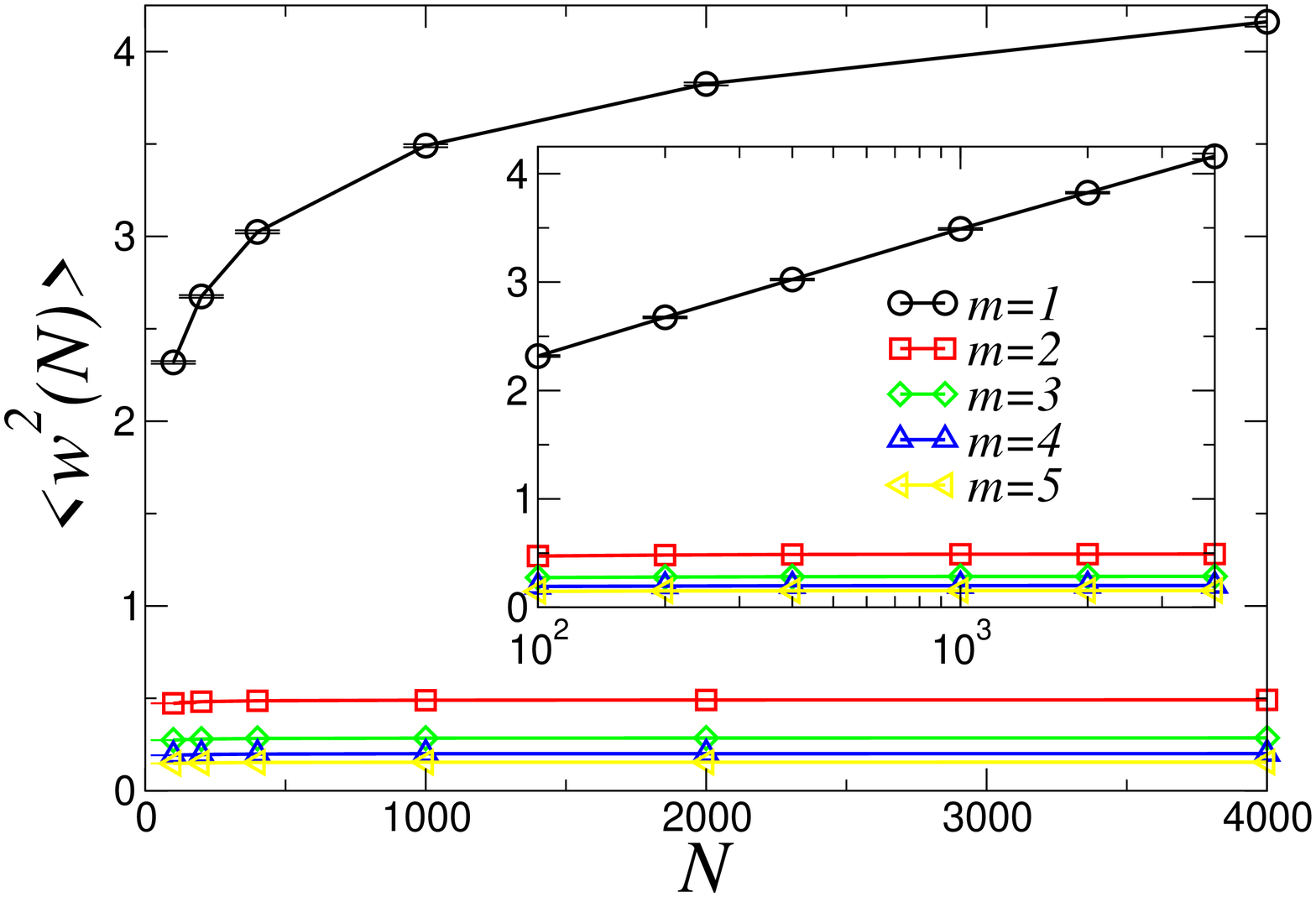} &
\hspace{-2mm}
\includegraphics[keepaspectratio=true,angle=0,width=60mm]
{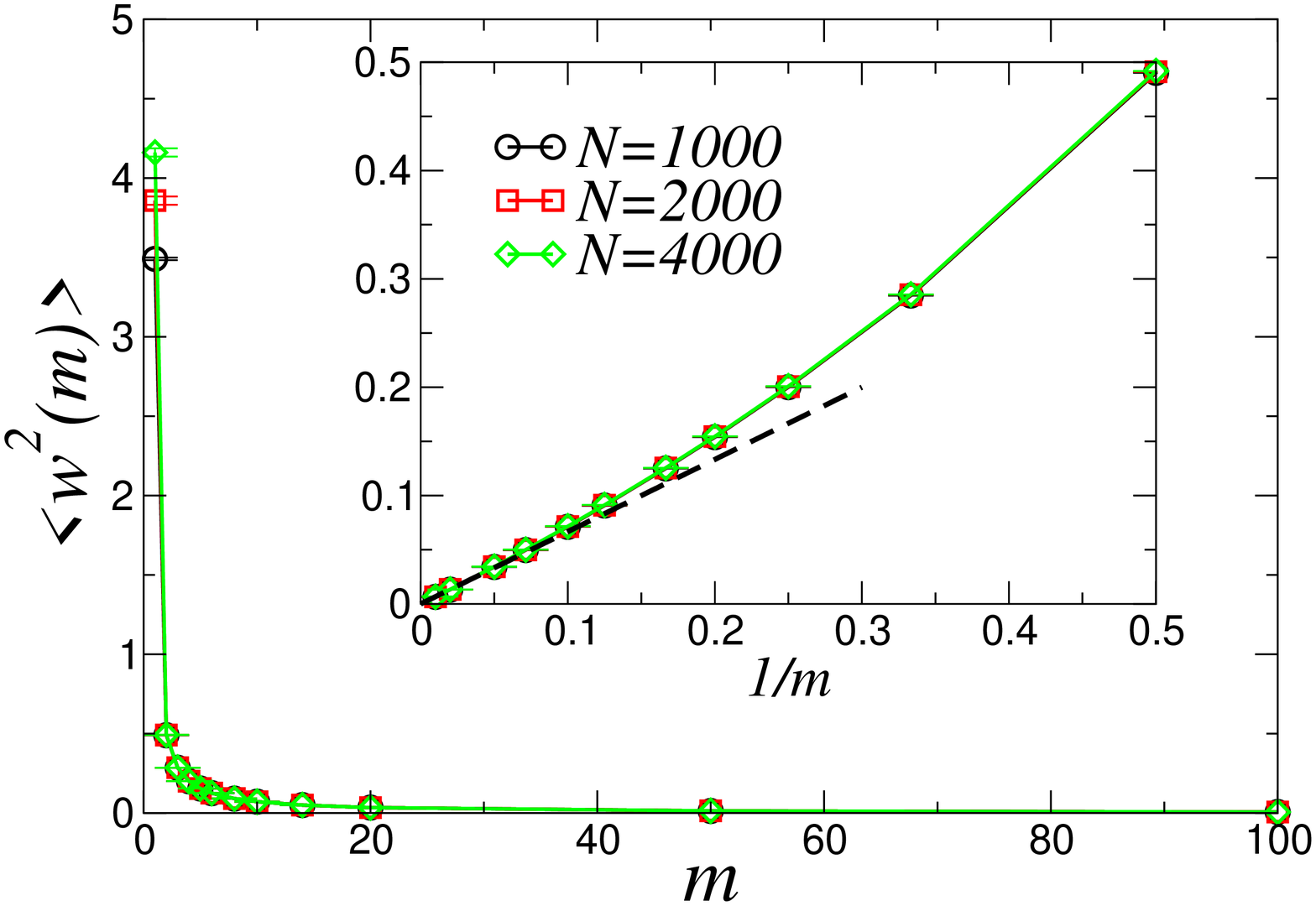} &
\hspace{-2mm}
\includegraphics[keepaspectratio=true,angle=0,width=60mm]
{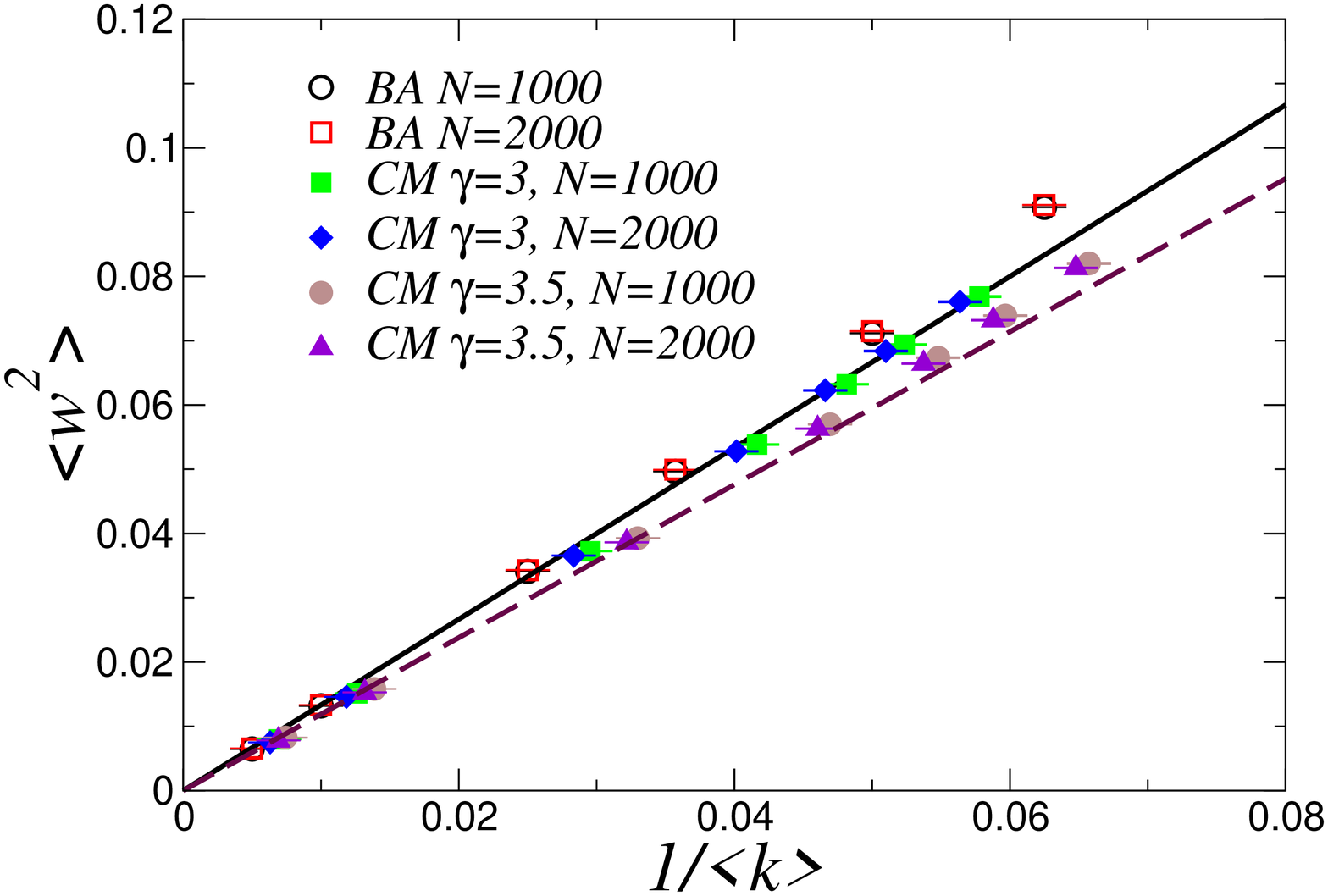}
\vspace{-0.3cm}
\\
\small{(a) Width vs system size} & \small{(b) Width vs m} &
\small{(c) Width vs $1/\langle k\rangle$}
\end{tabular}
\end{center}
\vspace{-0.5cm}
\caption{Steady-state width of the EW synchronization landscape
from exact numerical diagonalization using Eq.~(\ref{w2_exact}).
(a) For the BA network, as a function of $N$ for various
values of the minimum degree $m$. The inset shows the same data on log-lin scales.
(b) For the BA network, as a function of $m$ for different system sizes $N$.
The inset shows the behavior of the width vs $1/m$; the solid straight line
represents the MF result Eq.~(\ref{w2_MF_final}).
(c) For the BA and CM networks (with $\gamma=3.0$ and $\gamma=3.5$)
as a function of $1/\langle k\rangle$, where $\langle k\rangle$ is the average degree, for two system sizes.
The bold straight solid and dashed lines correspond to the MF result Eq.~(\ref{w2_MF_final})
with $\gamma=3.0$ and $\gamma=3.5$, respectively.}
\label{fig_exact_w2}
\end{figure*}

We also note that the average width, in principle, can also be
obtained by employing the density of states (dos) $\rho(\lambda)$
of the underlying network Laplacian through
$\langle w^2 \rangle = (1/N)\sum_{l=1}^{N-1} 1/\lambda_{l} \simeq
\int(1/\lambda)\rho(\lambda)d\lambda$,
in the asymptotic large-$N$ limit \cite{KOZMA03,KK_UGA}. Obtaining
the dos analytically, however, is a rather challenging task. Just
recently, using the replica method \cite{Bray1988,Rodgers2005},
Kim and Kahng obtained the dos for the Laplacian of SF graphs \cite{Kim2006},
which they were able to evaluate in the
asymptotic $1$$<<$$\langle k\rangle$$<<$$N$ limit. Utilizing their
result, we have checked and found full agreement
with Eq.~(\ref{w2_MF_final}) for the width.

While our MF approach above does not directly address the typical
size of the maximum relative height of the task-completion
landscape, the finding that the width is finite in the
$N\to\infty$ limit suggests that correlations between height
fluctuations at different nodes are weak (with the exception of the $m=1$ BA
tree). Then, one can argue on the scaling of the extremes as
follows. The largest fluctuations will most likely emerge from the
nodes with (or close to) the smallest degree $m$
[Eq.~(\ref{w_i})]. The typical size of the fluctuations on such
nodes, according Eq.~(\ref{w_i}), is
$\sqrt{\langle(\tau_i-\bar{\tau})^2\rangle}_{k_{i}=m}\sim
m^{-1/2}$. The expected number of nodes with the smallest degree $m$ is
$N_m\sim NP(m)\sim(\gamma-1)(N/m)$. Thus, assuming that the
fluctuations on these nodes are independent, we expect
\begin{equation}
\langle\Delta_{\max}\rangle =\langle\tau_{\max}-\bar{\tau}\rangle
\sim \sqrt{\langle(\tau_i-\bar{\tau})^2\rangle}_{k_{i}=m}
\left[\ln[(\gamma-1)(N/m)]\right]^{1/\delta}
\sim
m^{-1/2}\left[\ln(N)\right]^{1/\delta}\;,
\label{w2_ext_SF}
\end{equation}
and the distribution of the extremes is governed by the Gumbel
distribution in the asymptotic large-$N$ limit. Note that the
exponent $\delta$ depends on the details of the noise (or local
stochastic task increments), e.g., $\delta$$=$$2$ for Gaussian-like
tails and $\delta$$=$$1$ for exponential tails. The $m^{-1/2}$
pre-factor is also specific to the linear EW coupling on the
network by virtue of Eq.~(\ref{w_i}), and we do not expect to be
generally applicable. We do expect, however, that the weak
logarithmic divergence with the system size,
Eq.~(\ref{w2_ext_SF}), governed by the Gumbel distribution, will
hold for the actual simulated task-completion landscape which evolves
according to the synchronization rule
Eq.~(\ref{revolution}), on SF networks.

\end{subsection}

\begin{subsection}{Simulation Results}

In this subsection we present detailed results and analysis of the
simulations of the exact task-completion rule
Eq.~(\ref{revolution}) on BA and CM networks.
We simulated the task-completion
system on these networks and measured the steady-state width Eq.~(\ref{width})
and maximum fluctuations over many different network realizations and
generated their distributions.

Fig.~\ref{fig_scaling-ba} shows the average maximum fluctuations and
the width as function of $m$ for different system sizes ranging
from 100 to 10,000. Each data point was obtained  by averaging over ten
different network realizations. As it can be seen from
Fig.~\ref{fig_scaling-ba}(a), $\langle \Delta_{\max} \rangle$ rapidly
approaches a system-size-dependent constant. For the $m=1$ BA
model, the network is a tree, and $\langle \Delta_{\max} \rangle$
is visibly larger than for higher values of $m$.
Figure~\ref{fig_scaling-ba}(b) contains the same data points
as Fig.~\ref{fig_scaling-ba}(a), but the data is plotted as function of the system size,
for different values of $m$. The average maximum fluctuations in Fig.~\ref{fig_scaling-ba}(b)
scale logarithmically (or diverge weakly) with the system size
for all values of $m$. Again, the $m=1$ case is different from others
in terms of the pre-factors, although they all exhibit logarithmic divergence.
Thus, for BA networks, we find
\begin{equation}
\langle \Delta_{\max} \rangle \sim \ln(N) \;.
\label{eq_dmax_logN}
\end{equation}
supporting the MF prediction Eq.~(\ref{w2_ext_SF}). The ``clean"
logarithmic scaling in Fig.~\ref{fig_scaling-ba}(b)
is the result of the individual task
distributions $P(\tau_i-\bar\tau)$ having an exponential tail ($\delta=1$) for the
task-completion rule.

With the exception of the $m$$=$$1$ BA tree, the width converges
to an essentially {\em system-size-independent}, but non-zero,
value [Fig.~\ref{fig_scaling-ba}(c)]. The main difference, with respect to the MF prediction, is
the non-vanishing ``intrinsic" width as $m\to\infty$. This behavior is
due to the specific synchronization rule
Eq.~(\ref{revolution}). Namely, when $m$ is large,
only a couple of nodes are allowed to increment,
hence the landscape fluctuations are essentially governed by the
stochastic task increments at these nodes. Since
the variance of the local fluctuations is unity, the width
of the landscape converges to unity for large $m$ [Fig.~\ref{fig_scaling-ba}(c)].
It can be seen in Fig.~\ref{fig_scaling-ba}(d) that the width has a logarithmic
divergence for $m=1$ and it is a constant for $m>1$, i.e., for
large $N$
\begin{equation}
\langle w^2(N) \rangle \sim
\left\{
\begin{array}{ll}
\ln(N),    & \text{if}~m=1\\
\text{const.},    & \text{if}~m>1\\
\end{array}
\right. \;.
\label{eq_w2_logN}
\end{equation}

\begin{figure*}
\begin{center}
\begin{tabular}{cc}
\includegraphics[keepaspectratio=true,angle=0,width=83mm]
{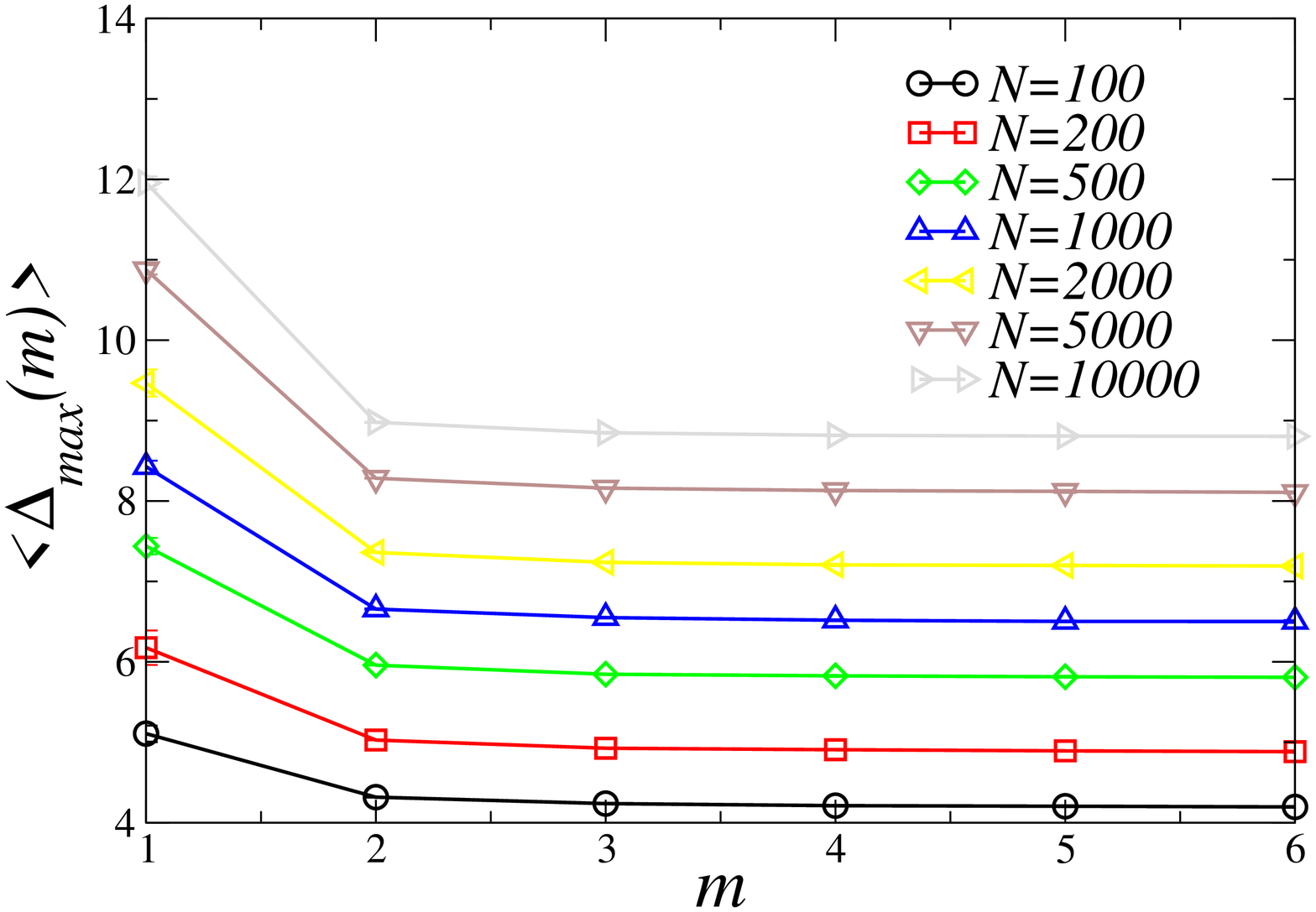} &
\hspace{-2mm}
\includegraphics[keepaspectratio=true,angle=0,width=83mm]
{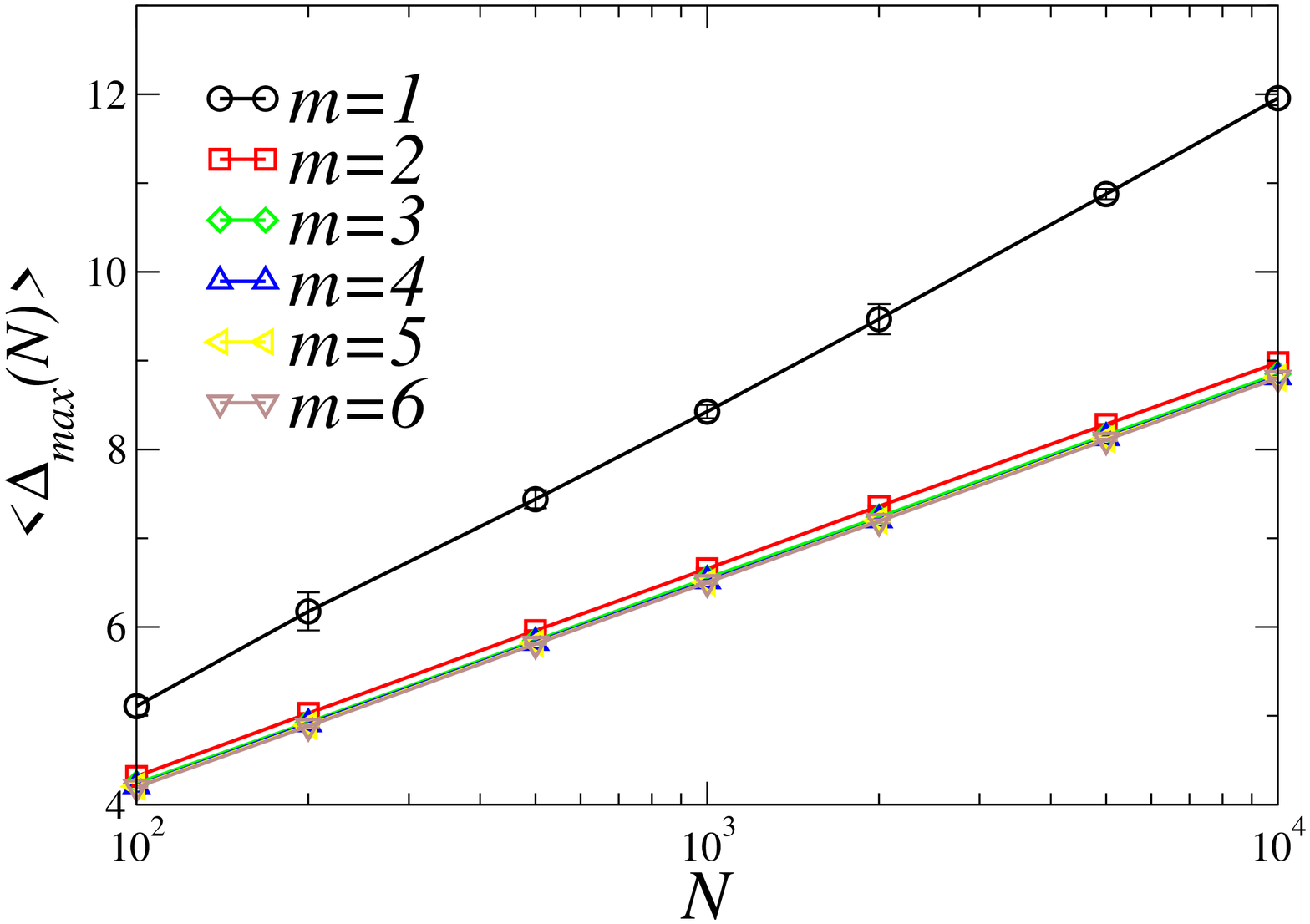}
\vspace{-0.3cm}
\\
\small{(a) Maximum fluctuations vs m} & \small{(b) Maximum fluctuations vs system size}
\\
\includegraphics[keepaspectratio=true,angle=0,width=83mm]
{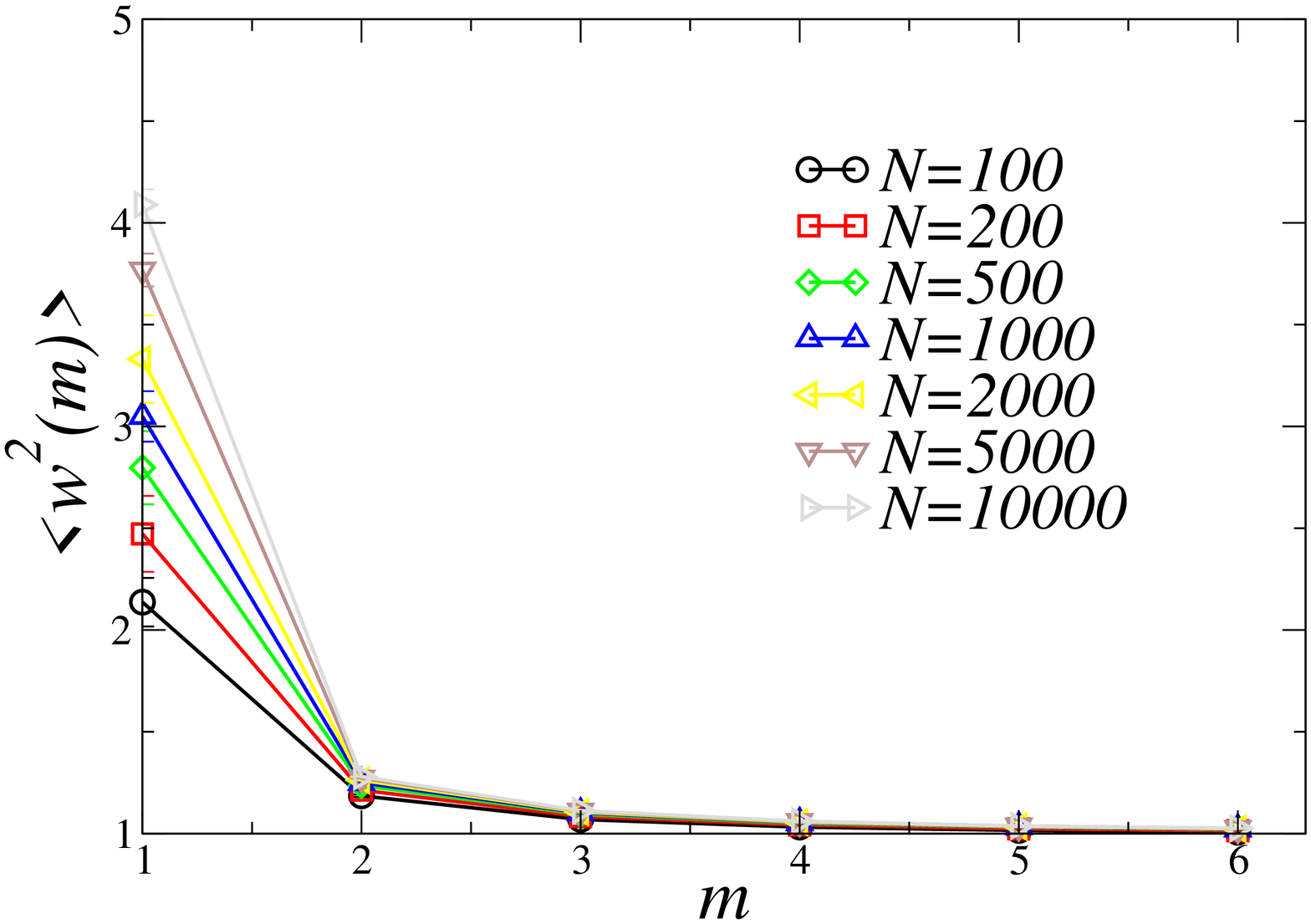} &
\hspace{-2mm}
\includegraphics[keepaspectratio=true,angle=0,width=83mm]
{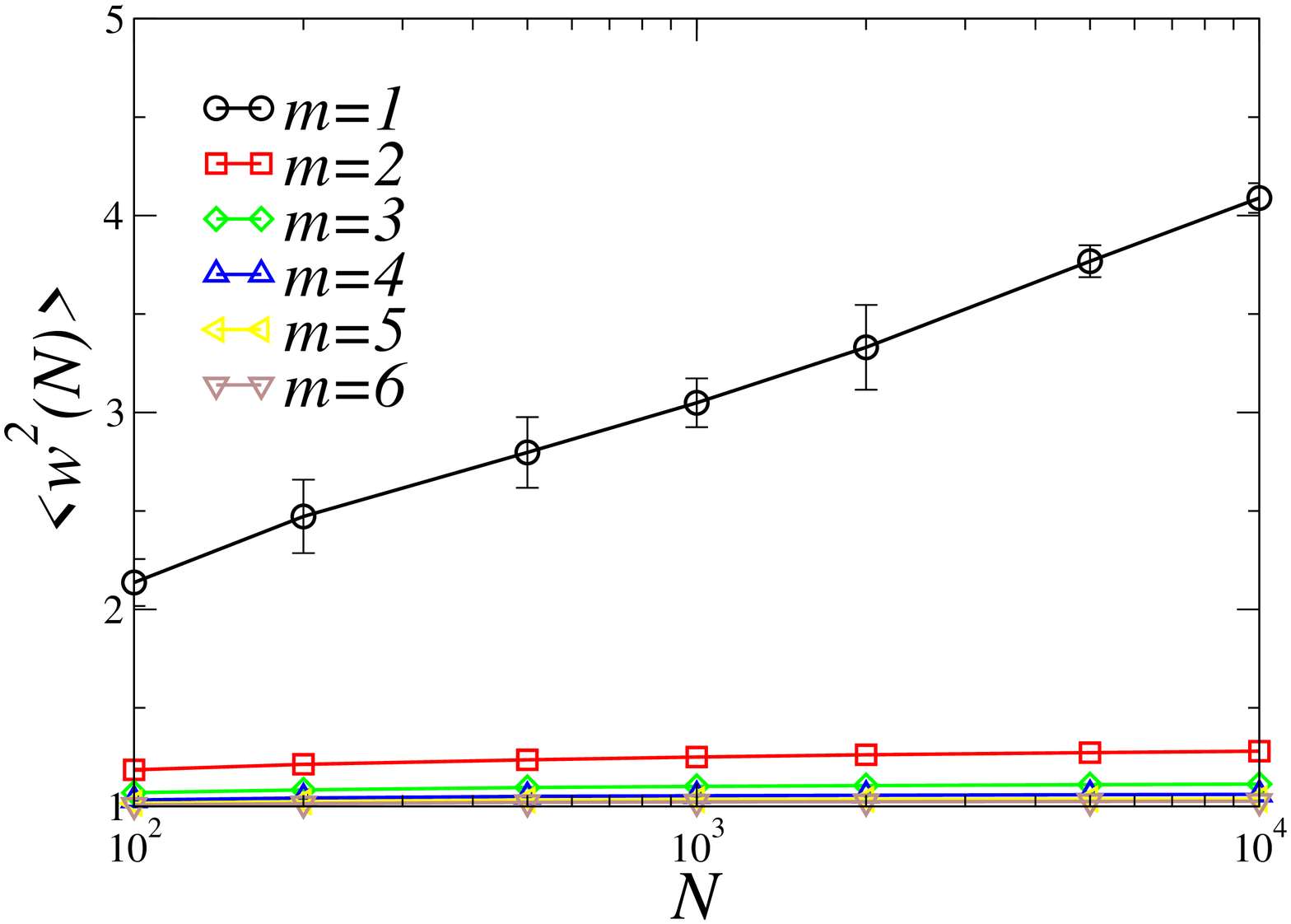}
\vspace{-0.3cm}
\\
\small{(c) Width vs m} & \small{(d) Width vs system size}
\end{tabular}
\end{center}
\vspace{-0.5cm}
\caption{Average maximum fluctuations and average width for SF networks
generated by the BA model. The data points are obtained by averaging over ten different network realizations.
}
\label{fig_scaling-ba}
\end{figure*}

The CM network has very similar characteristics to the BA model in terms of
the scaling of maximum fluctuations and width.
Fig.~\ref{fig_scaling-mr}(a) shows that the average maximum
fluctuations for CM network with $\gamma=3$, as a function of $m$, have the same behavior as for the BA model. Since the
CM generates a (single-component) connected network with very low probability for $m=1$,
we only present results for $m>1$. As it can be seen in Fig.~\ref{fig_scaling-mr}(b),
$\langle \Delta_{\max} \rangle$ increases logarithmically with the
system size. The data points were obtained by averaging over ten
different realizations of the network. One observes that at low values of
$\gamma$, $\langle \Delta_{\max}(m=2) \rangle$ is
closer to $\langle \Delta_{\max}(m>2) \rangle$ and the difference
increases as $\gamma$ increases. This implies that having fewer
high-degree nodes in the network ($\gamma=4$) separates $\langle
\Delta_{\max}(m=2) \rangle$ from $\langle \Delta_{\max}(m>2) \rangle$.

\begin{figure*}
\begin{center}
\begin{tabular}{cc}
\includegraphics[keepaspectratio=true,angle=0,width=83mm]
{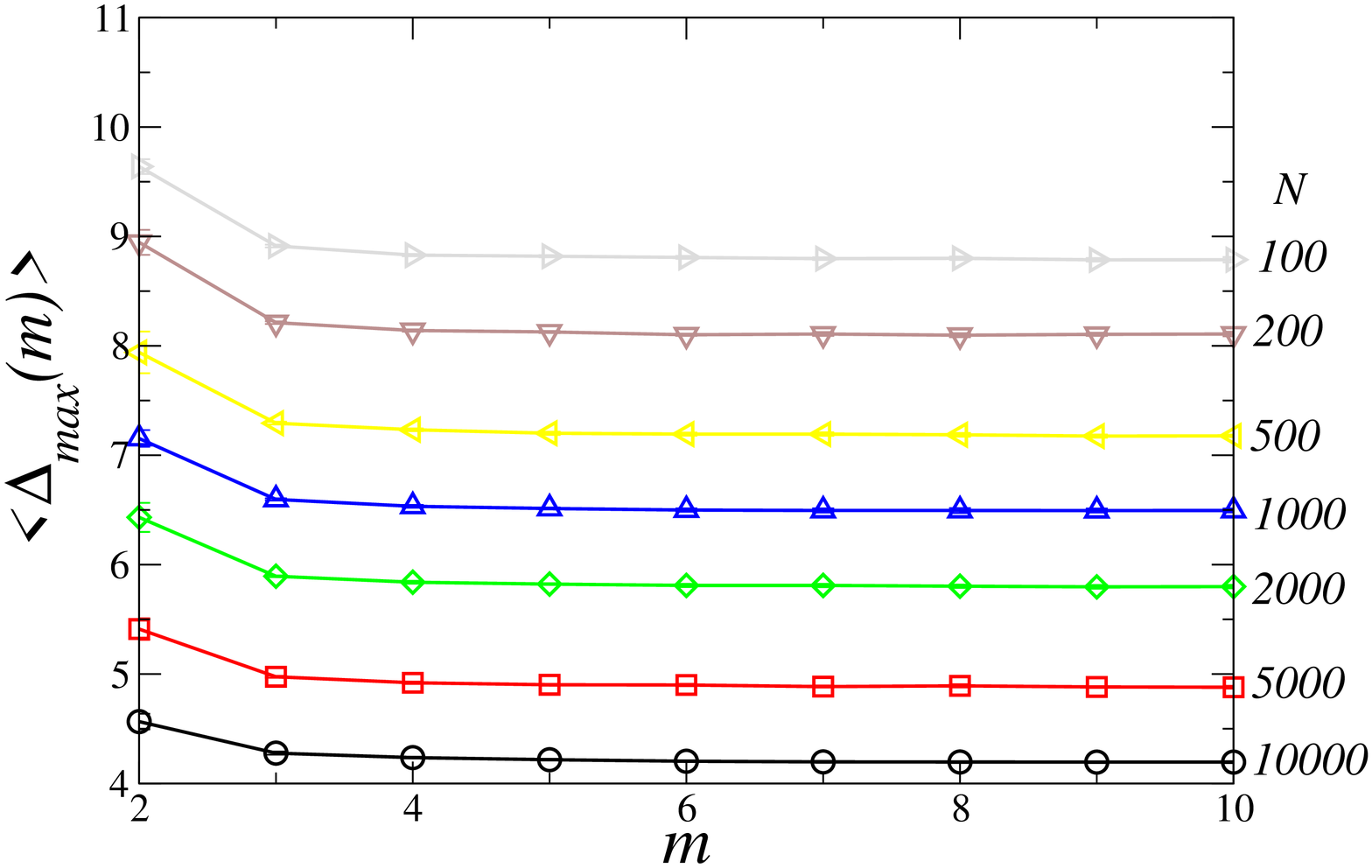} &
\hspace{-2mm}
\includegraphics[keepaspectratio=true,angle=0,width=83mm]
{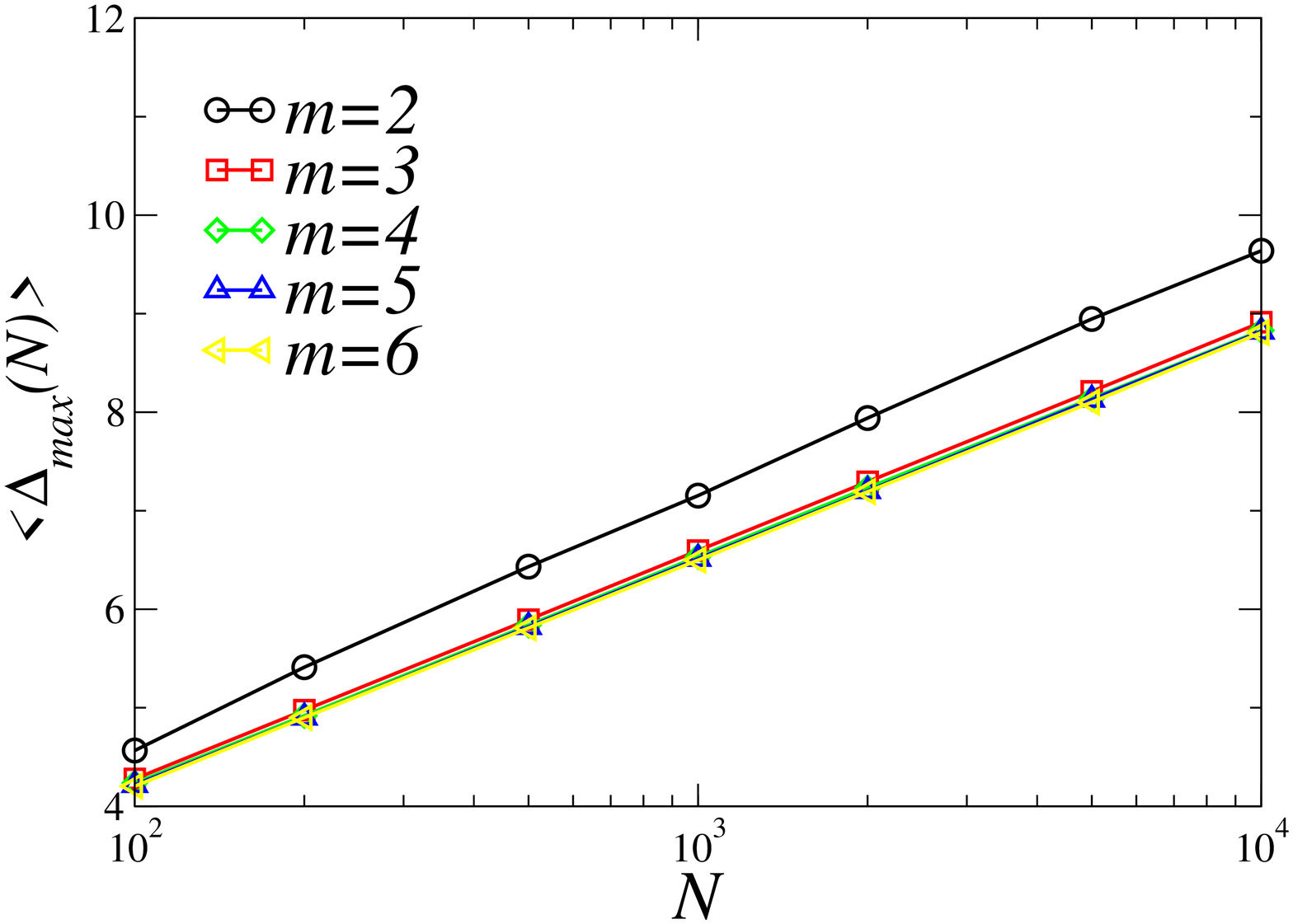}
\vspace{-0.3cm}
\\
\small{(a) Maximum fluctuations vs $m$} & \small{(b) Maximum fluctuations vs system size}
\\
\includegraphics[keepaspectratio=true,angle=0,width=83mm]
{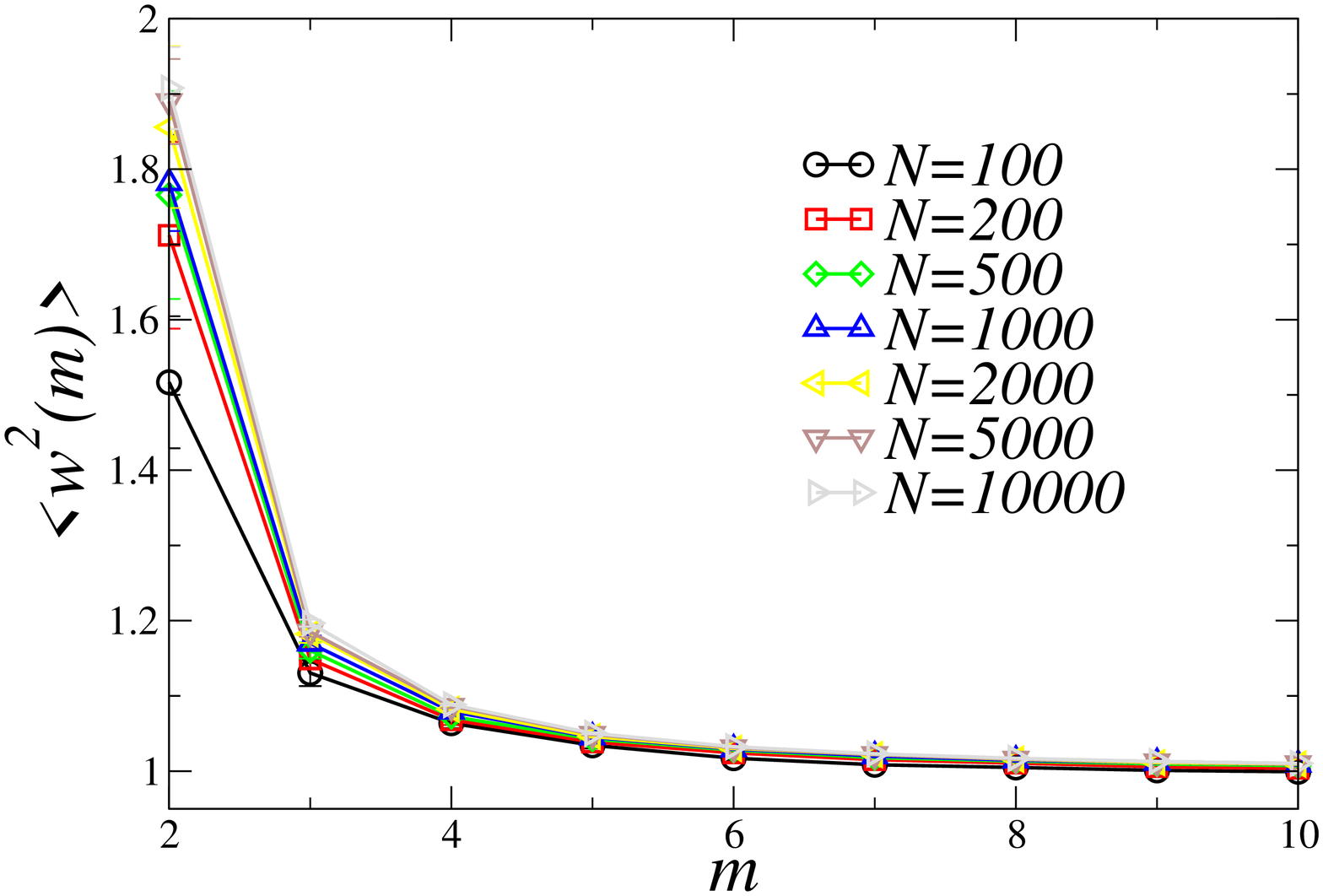} &
\hspace{-2mm}
\includegraphics[keepaspectratio=true,angle=0,width=83mm]
{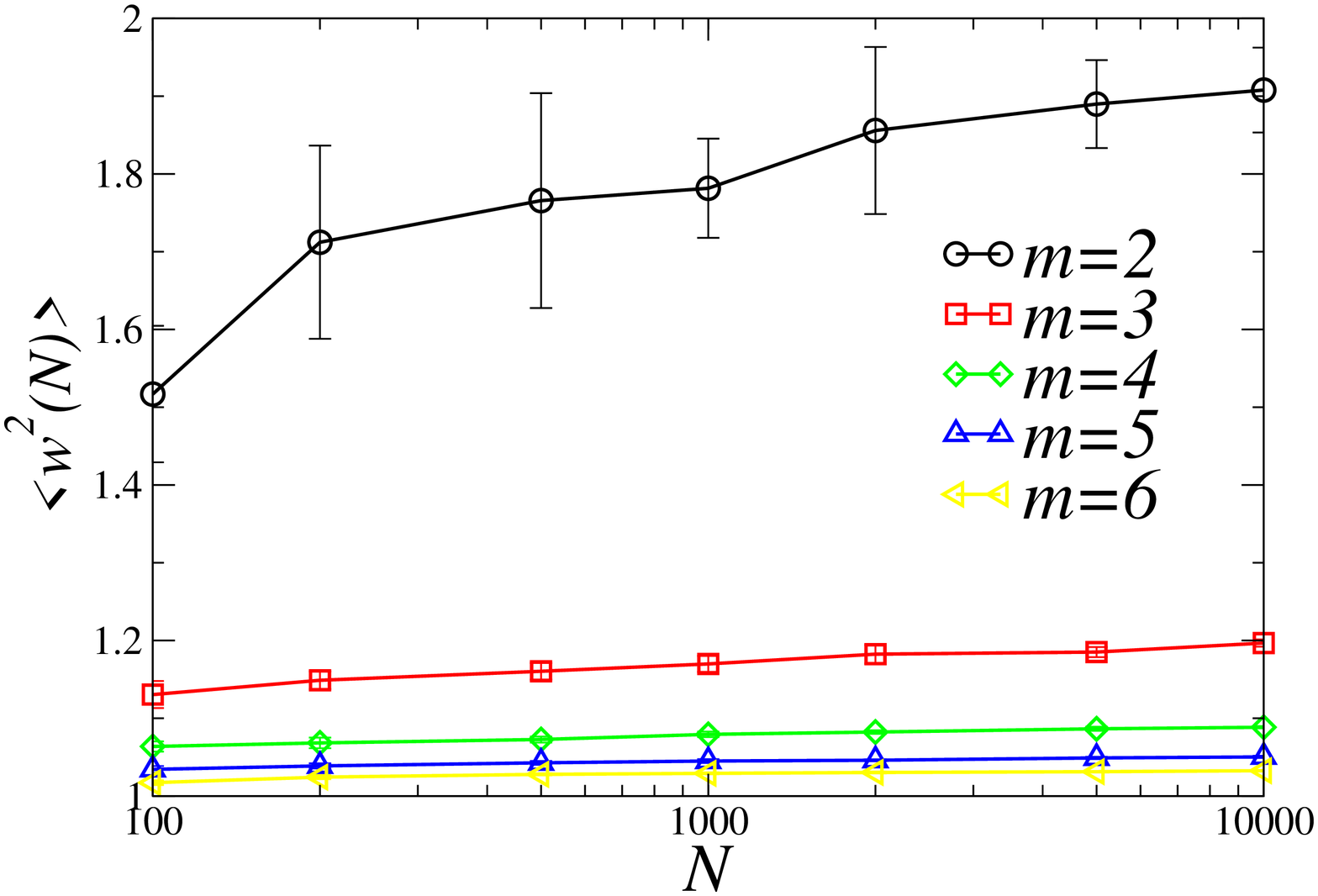}
\vspace{-0.3cm}
\\
\small{(c) Width vs m} & \small{(d) Width vs system size}
\vspace{-0.5cm}
\end{tabular}
\end{center}
\caption{Average maximum fluctuations and the width for SF CM networks with $\gamma=3$ versus system size and
$m$. The data points are obtained by averaging over ten different network realizations.
}
\label{fig_scaling-mr}
\end{figure*}

The average width as a function of $m$ [Fig.~\ref{fig_scaling-mr}(c)] for $\gamma=3$
decays and reaches its asymptotic value.
The scaling of the width as a function of the system size is
in Fig.~\ref{fig_scaling-mr}(d). The error bars are
quite visible although we use at least ten different network
realizations. There is a slight increase in the width as a function of the
system size for $m=2$, whereas for larger $m$,
the width quickly saturates as a function of $N$.

Fig.~\ref{fig_dist-ba}(a) shows the individual task
distributions (parent distributions for the extremes) for the BA
network with $m=2$ and $N=10^4$. These distributions exhibit simple exponential tails
($\delta=1$) (inherited from the local exponential
task-increments). The nodes for which the distributions presented in Fig.~\ref{fig_dist-ba}(a) were
selected manually according to their degrees. We selected high
(also maximum), middle and low (also minimum) degree nodes.
The legend shows both the index of the node, which is
the ``age'' of the node according to the preferential attachment
procedure in the BA model, and its degree. For larger $m$ the
distributions yield better collapse to an exponential, and also
their negative parts (for the fluctuations below the mean) become smaller, i.e., the
fluctuations are asymmetric about the mean (due to the specific local task-increment rules).
It can also be seen that the negative part of the individual task distributions are not
pure exponentials, i.e., $\delta>1$, which makes the convergence
of the fluctuations of the minima ($\Delta_{min}$) toward their
limiting Gumbel distribution much slower.

Figs.~\ref{fig_dist-ba}(b) and \ref{fig_dist-ba}(c) show
the distributions of the maximum fluctuations and the width
for the BA model with $m=2$ and their comparison to the Gumbel and Gaussian distributions, respectively.
The insets in Figs.~\ref{fig_dist-ba}(b) and \ref{fig_dist-ba}(c)
have the same data as the main graph but scaled to
zero mean and unit variance, and in a log-linear scale to show the collapse to
the limit distributions in the tails. The pure exponential behavior of individual
task distributions in Fig.~\ref{fig_dist-ba}(a) suggests that the limit distribution
for the maximum fluctuations would be a Gumbel
distribution. As it can be seen in Fig.~\ref{fig_dist-ba}(b)
the limit distributions have better collapse as the system size gets larger.
The width distributions for the BA model with $m=2$
are plotted in Fig.~\ref{fig_dist-ba}(c). The mean-field approximation predicts that
the local task fluctuations are decoupled and consequently the distribution of the width
converges to a Gaussian for large enough systems. We verified this prediction
and showed that the width distributions converge to delta
functions, and when they are scaled to zero mean and unit variance they collapse
to a standard Gaussian distribution for large enough system size.
When $m=1$ the width distribution converges to a nontrivial shape with an exponential
tail.
\begin{figure*}
\begin{center}
\begin{tabular}{ccc}
\includegraphics[keepaspectratio=true,angle=0,width=60mm]
{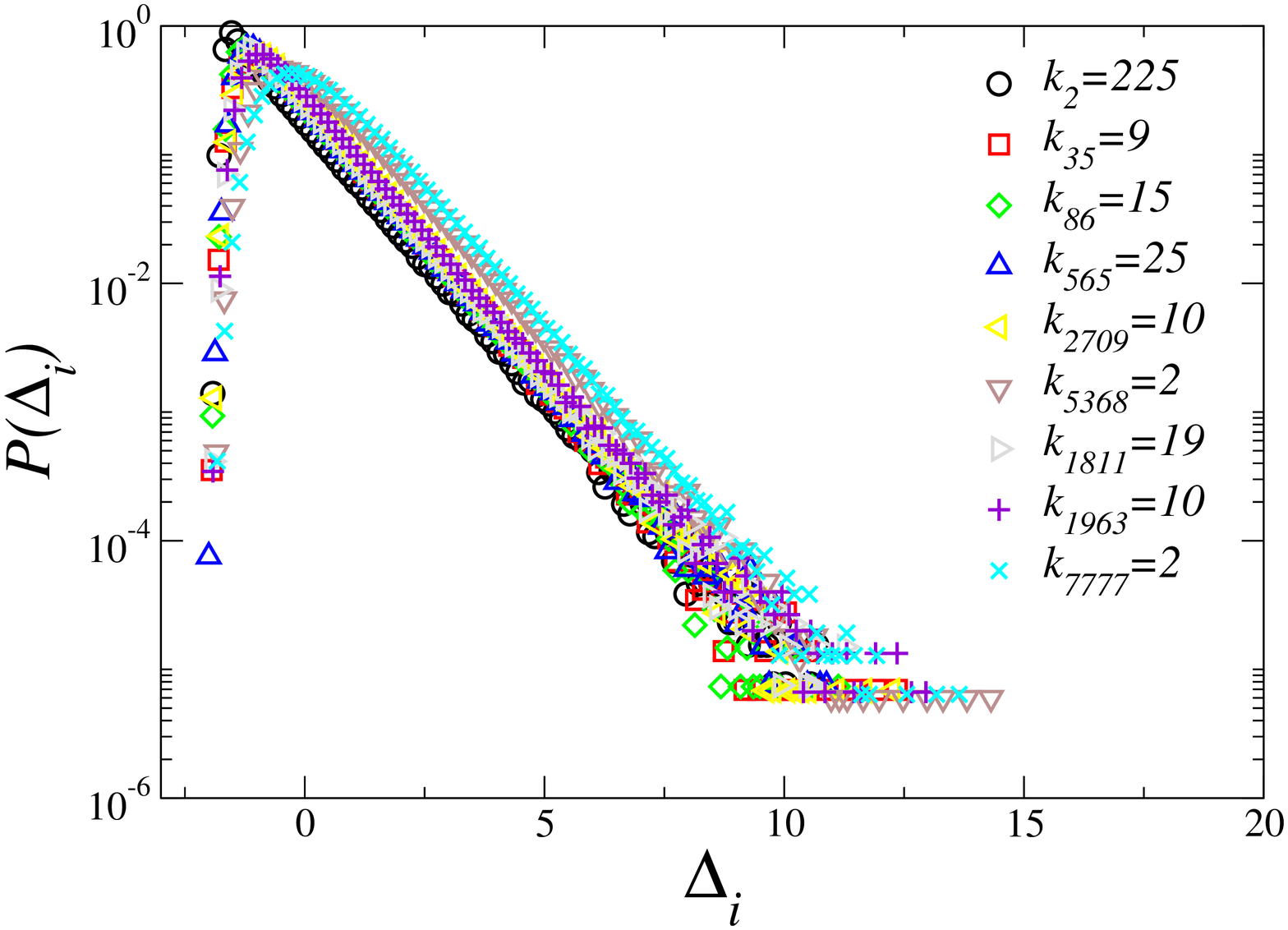} &
\hspace{-2mm}
\includegraphics[keepaspectratio=true,angle=0,width=60mm]
{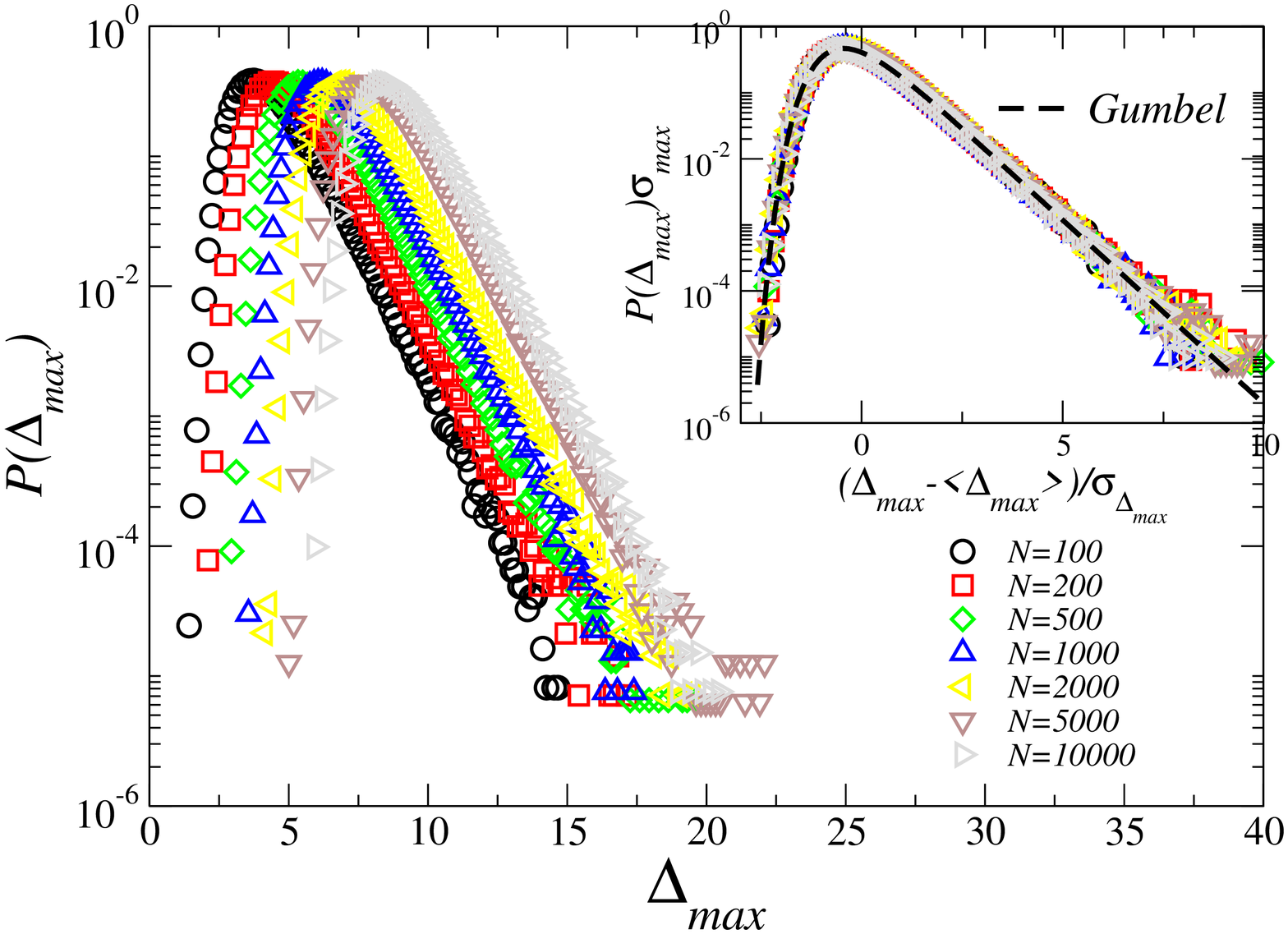} &
\hspace{-2mm}
\includegraphics[keepaspectratio=true,angle=0,width=60mm]
{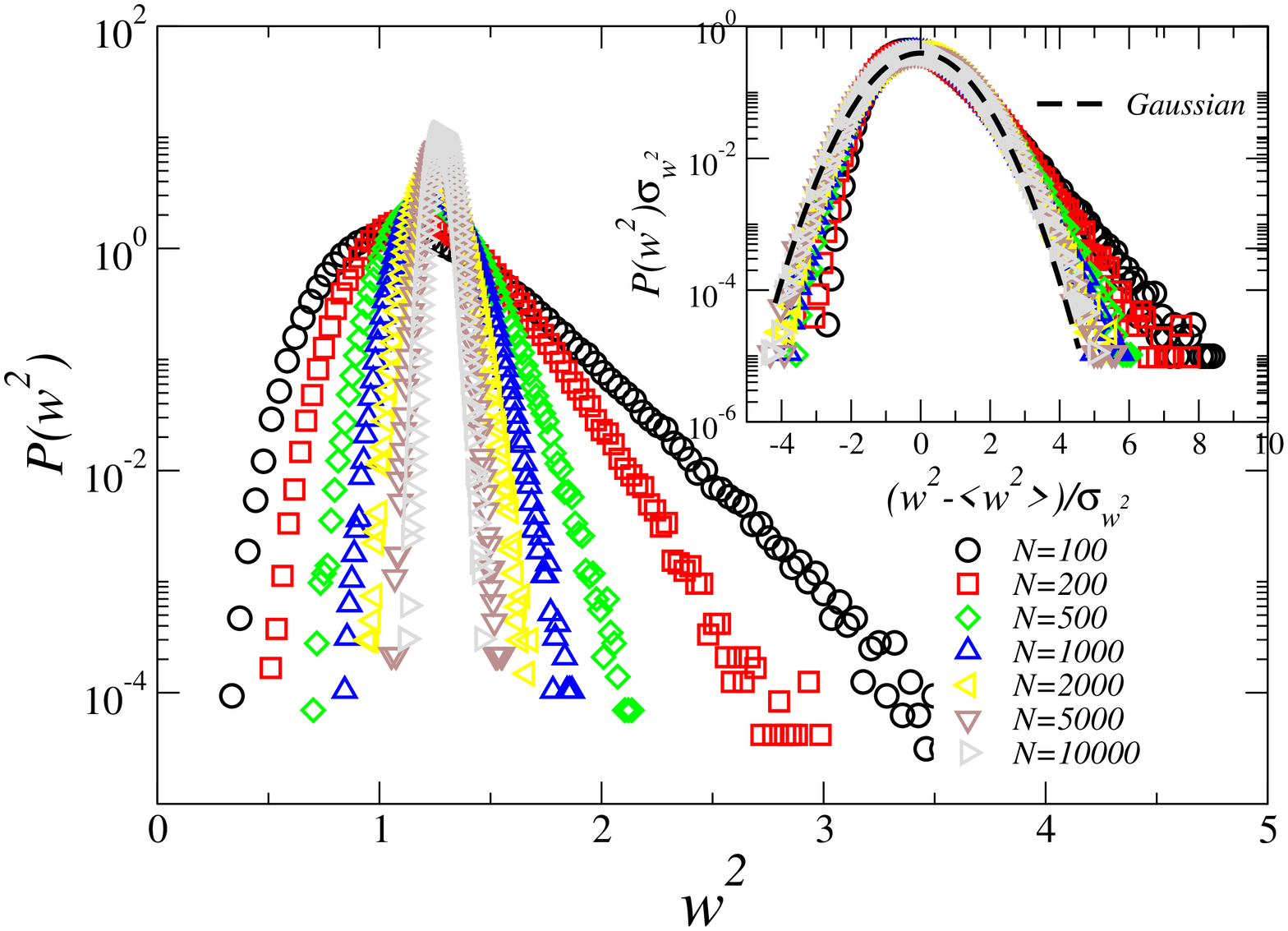}
\vspace{-0.3cm}
\\
\small{(a) Individual fluctuations distribution} & \small{(b) Maximum fluctuations distribution} &
\small{(c) Width distribution}
\vspace{-0.5cm}
\end{tabular}
\end{center}
\caption{Distribution of individual fluctuations (a), maximum fluctuations
(b) and the width (c) for the BA model with $m=2$. The different individual
fluctuations distributions in (a) are for different degree values ranging
from the maximum to the minimum. The inset in (b) shows the maximum
fluctuations distribution scaled to zero mean and unit
variance in a log-linear scale. The dashed curves in the insets represent the Gumbel
pdf [Eq.~(\ref{gumbel_pdf_scaled})] in (b)
and Gaussian pdf in (c) scaled in the same way. The system size is $N=10^4$.}
\label{fig_dist-ba}
\end{figure*}

Similar to individual task distributions in the BA model, the CM has pure exponential
distributions in the tail as shown in Fig.~\ref{fig_dist-mr}(a) for $m=3$, $\gamma=3$,
and system size $N=10^4$. The nodes are selected according to their degrees,
i.e., a few high, middle and low-degree nodes. The maximum fluctuation distributions
converge to Gumbel distributions even for small systems
as it can be seen in Fig.~\ref{fig_dist-mr}(b).
It can be concluded for the CM that the width distributions
converge to delta functions as system size goes to infinity and when they are scaled to zero mean and unit variance
they converge to the standard Gaussian [Fig.~\ref{fig_dist-mr}(c)].
For the somewhat subtle case of the CM network with $m=2$ the convergence to a finite width
is slow [see Fig.~\ref{fig_scaling-mr}(d)] possibly due to strong finite system-size effects.

\begin{figure*}
\begin{center}
\begin{tabular}{ccc}
\includegraphics[keepaspectratio=true,angle=0,width=60mm]
{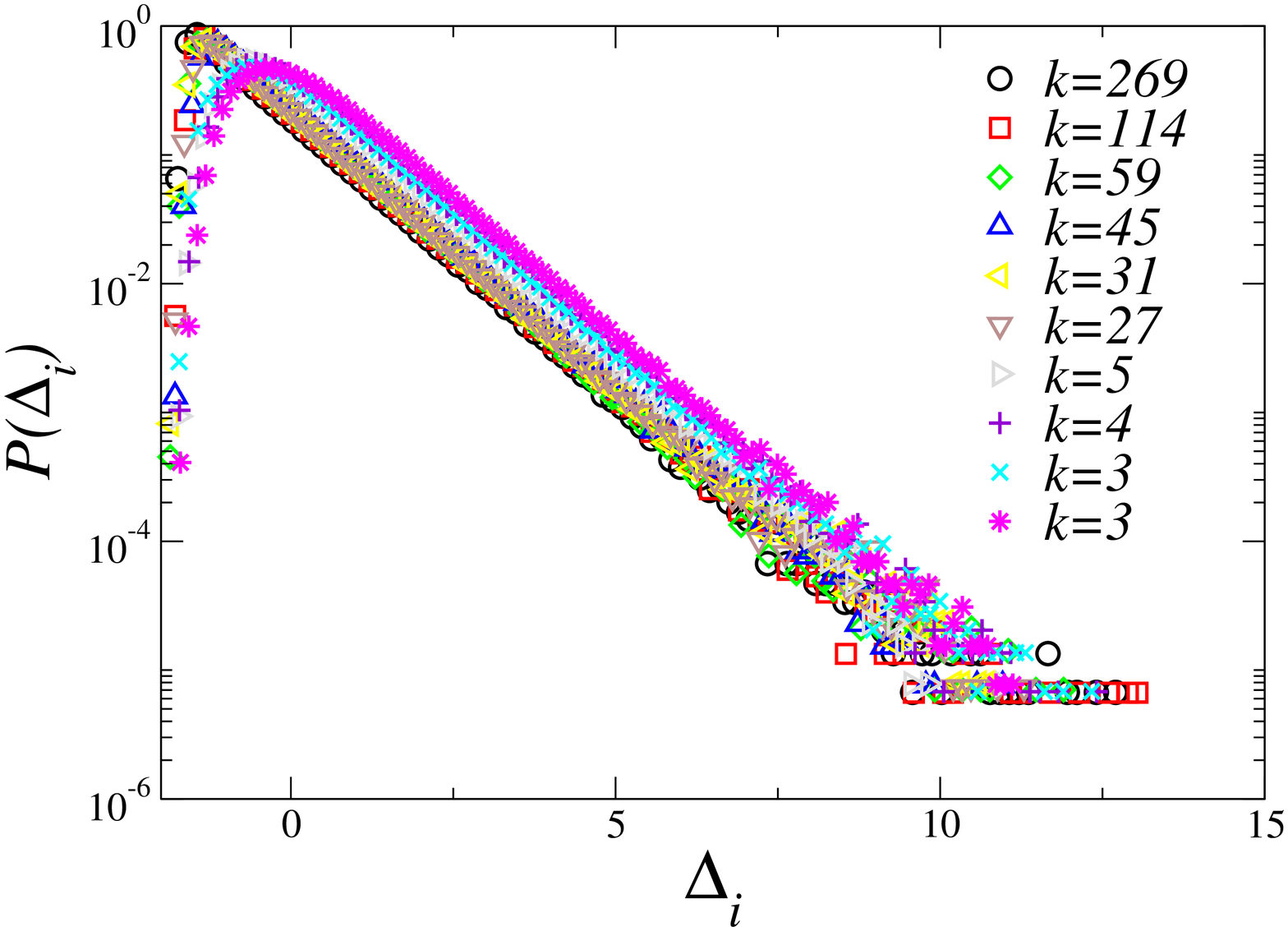} &
\hspace{-2mm}
\includegraphics[keepaspectratio=true,angle=0,width=60mm]
{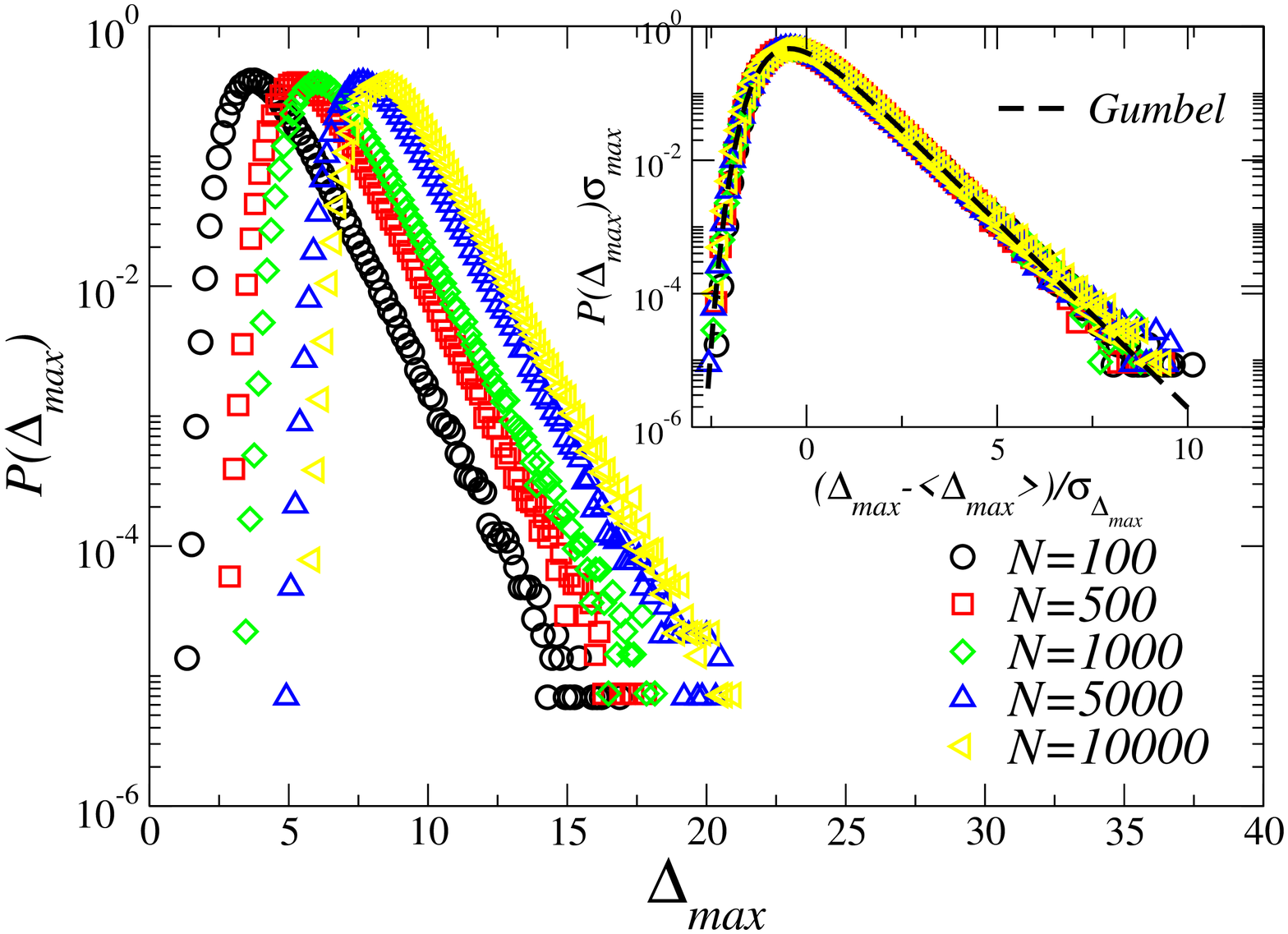} &
\hspace{-2mm}
\includegraphics[keepaspectratio=true,angle=0,width=60mm]
{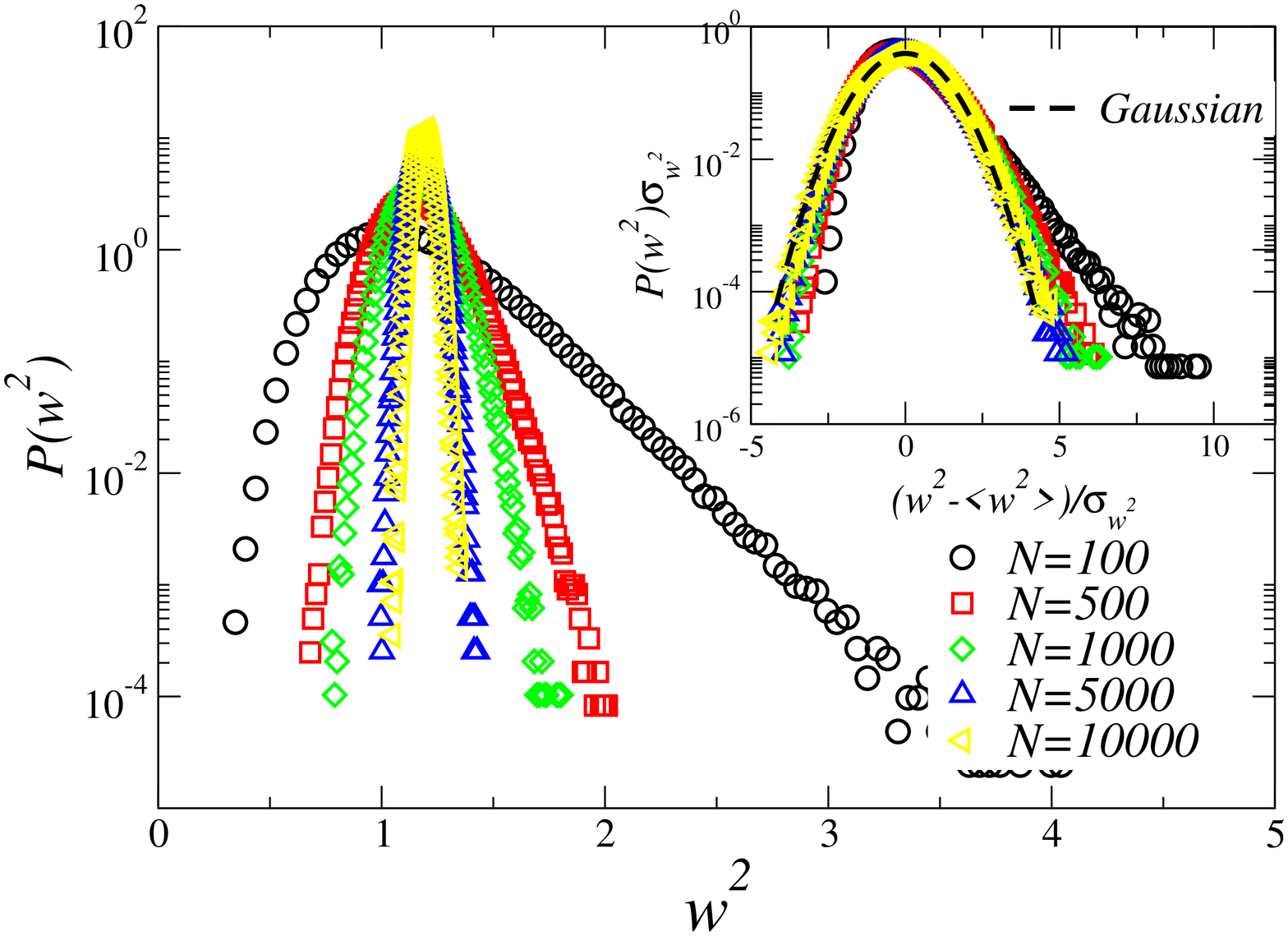}
\vspace{-0.3cm}
\\
\small{(a) Individual fluctuations distribution} & \small{(b) Maximum fluctuations distribution} &
\small{(c) Width distribution}
\vspace{-0.5cm}
\end{tabular}
\end{center}
\caption{Same as Fig.~\ref{fig_dist-ba} for the CM network with $\gamma=3$ and $m=3$.}
\label{fig_dist-mr}
\end{figure*}

\end{subsection}
\end{section}

\begin{section}{Summary and Conclusions}

In summary, we considered the average and maximum fluctuations
above the mean in scale-free task-completion landscapes with
local relaxation, unbounded local variables, and short-tailed
noise. We argued, that when the interaction topology is scale-free,
having a power-law degree distribution, the statistics of the
extremes is governed by the Gumbel distribution while the
distribution of average fluctuations converges to a Gaussian when
appropriately scaled. This finding directly addresses
synchronizability in generic task-completion systems with
scale-free network topology where relaxation through the links is
the relevant node-to-node process and effectively governs the
dynamics. Analogous questions for heavy-tailed noise distribution
on complex networks have relevance to various transport phenomena
in natural, artificial, and social systems
\cite{flux,BARA04,CROVELLA97,CROVELLA98,LELAND94,CSABAI94,PAXSON95}.
For example, ``bursty" temporal processes in queuing networks
have been recently attributed to online activities initiated by
humans \cite{BARA05}. Correspondingly, one shall then
study extreme fluctuations in task-completion landscapes where the
local task increments are power-law distributed. Heavy-tailed
noise typically generates similarly tailed local field variables
through the collective dynamics in SW
\cite{GUCLU_CNLS,GUCLU05_FNL} and SF networks \cite{GUCLU_FUTURE}.
Then, the largest fluctuations will likely diverge as a power law
with the system size, expectedly governed by the Fr\'echet distribution
\cite{GUMBEL,GALAMBOS}.

\end{section}

\begin{acknowledgments}

We thank S. Majumdar for providing us with the numerically
evaluated Airy distribution \cite{MAJUMDAR04,MAJUMDAR05}, shown in
Fig.~\ref{fig_dist-regular}(b) for comparison.
We also thank Z. R\'acz for valuable discussions. H.G. was supported
by the U.S. DOE through DE-AC52-06NA25396. G.K. acknowledges the
financial support of NSF through DMR-0426488 and RPI's Seed Grant.
Z.T. has been supported by the University of Notre Dame.

\end{acknowledgments}

\end{document}